\documentclass{article}
\usepackage[UTF8, scheme=plain, punct=plain,]{ctex}% 删除 zihao=false
\usepackage{amsmath}      %数学公式
\usepackage{amsfonts}     %数学字体
\usepackage{mathrsfs}
\usepackage{cases}%20250228

\usepackage{adjustbox} % 在导言区

\usepackage[numbers]{natbib}
\usepackage{graphicx}
\usepackage{color}
\usepackage{caption} %20250307slx
\usepackage{longtable} 

\usepackage{enumerate} %20240322slx编号格式-enumerate
 
\usepackage{ulem}    %20240305slx \sout斜杆
\usepackage{cancel} % 20240325slx x斜杠划去
\usepackage{pst-tree}
\usepackage{forest}
\usepackage{qtree}

\usepackage{sgamex}
\usepackage{egameps}
\usepackage[most]{tcolorbox}

\usepackage{tabularx}
\usepackage{subcaption}

\usepackage{multirow}
\usepackage{booktabs}
\usepackage{booktabs}
\usepackage{geometry}
\usepackage{url}
\usepackage{istgame}
\usepackage{tabu,makecell,multirow}
\usepackage{hyperref}  % 最后加载
\hypersetup{
  hidelinks,
  colorlinks=true,
  allcolors=black,
  pdfstartview=Fit,
  breaklinks=true
}

\usepackage{makeidx}
\makeindex
\DeclareExpandableDocumentCommand\xcol{mO{c}m}{\multicolumn{#1}{#2}{#3#3}}

\usepackage{listings}  %20240326slx添加python代码
\usepackage{xcolor}

\definecolor{codegreen}{rgb}{0,0.6,0}  
\definecolor{codegray}{rgb}{0.5,0.5,0.5}  
\definecolor{codepurple}{rgb}{0.58,0,0.82}  
\definecolor{backcolour}{rgb}{0.95,0.95,0.92}  
  
\lstdefinestyle{mystyle}{  
    backgroundcolor=\color{backcolour},     
    commentstyle=\color{codegreen},  
    keywordstyle=\color{magenta},  
    numberstyle=\tiny\color{codegray},  
    stringstyle=\color{codepurple},  
    basicstyle=\footnotesize\footnotesize,  
    breakatwhitespace=false,           
    breaklines=true,                   
    captionpos=b,                      
    keepspaces=true,                   
    numbers=left,                      
    numbersep=5pt,                    
    showspaces=false,                  
    showstringspaces=false,  
    showtabs=false,                    
    tabsize=2  
}

\title{Eigenmanifold in Game: Evidence from human continuous strategy game experiments\footnote{We thank Daniel Friedman for his helpful comments, especially for his supporting on our criticism on the ``First a hammer east, then a stick west'' condition of the research field in experimental investigations of evolutionary game theory. Corresponding author: wangzj@zju.edu.cn}}
\author{Wang Zhijian, Yao Qinmei, Shan Lixia, Wang Yijia, Zhang Chuchen  \\ Experimental Social Science Laboratory, Zhejiang University, China}
\date{\today}

\begin{document}
\maketitle
\tableofcontents
\newpage

\section*{Abstract\label{sec:abstract}}

In evolutionary game dynamics, there exists a hypothesis, which states that, the dynamic structure of the game's steady-state system is characterized by the linear superposition of eigenmanifolds, which depends specifically on the complex eigenvector structure at the Nash equilibrium and is ultimately governed by the game dynamics equations.  
This hypothesis has been supported widely in discrete strategy games. In continuous-strategy game, using experimental data from human-subject games, this paper finds that the hypothesis is {also significantly supported} A simplified model is provided to capture the observed wave mechanics of the dominant eigenmanifold, which is the {algebraic representation} of the complex  {eigenvector},  generated by the utility-driven evolutionary game dynamics.

\section{Introduction\label{sec:Intro}}

\subsection{Research question\label{sec:Research_question}}

 In game theory, the question that needs to be answered after Nash equilibrium is: how does the game move? \cite{nash1951noncooperative,friedman1998evolutionary,fudenberg1998theory,Newton2018,burnham2015experimental}  
The following is an analogy. When introducing a new substance, the first question a natural scientist needs to answer is its composition; after answering the composition of the new substance, the next question to be answered is its molecular structure. Faced with a new game, the first question a social scientist needs to answer is where its (Nash) equilibrium lies in the strategy space. Similarly, after understanding the equilibrium, the next question is to answered the movement  structure\cite{Newton2018,burnham2015experimental,duffy2022learning} . \\

Movement structure have two classes: time and space. In time, converge speed and frequency of cycle are the observation \cite{chen2004when,dan2014,wang2014}; In space, where to convergence and what evolution patterns are the observation. 
 {Recently, the two-dimensional (2D) beauty game is a typical study of movement in 2D space, answering where the dynamics goes and how it moves towards the end point (Nash equilibrium) of the game }\cite{duffy2022learning}. \\

But some game never get the end point, for example, the social cycle in rock-paper-scissors \cite{2013cycle,dan2014,2015nowak,2011coyness,wang2014social,wang2017}, which 
is a movement structure of cyclic pattern in space.  {Cycles are the main concern of this study}. Recently, in discrete strategy games, the dynamic manifold structure (cycle) is clear \cite{wang2017,Wang2022A4,wang2014,wang2014social,WY2020,2021Qinmei}; however, the representation theory and experimental measurement of the dynamic manifold structure of continuous strategy space game systems remain challenging \cite{dan2021price,dan2026Cyclical}. {This gap has persisted for decades} \cite{dan1984informational,dan2003,friedman1998evolutionary}

In high-dimensional discrete strategy game dynamics, we introduce a eigenmanifold \footnote{In this paper, the term 'eigenmanifold' (or more precisely, the eigenmanifold vector) refers to the vector $\sigma_k$, whose elements are real-valued scalars computed from the components of the eigenvector $\eta_k$. Its components depend solely on the magnitudes and phases of $\eta_k$, and are independent of the corresponding eigenvalue $\lambda_k$. We adopt this terminology because $\sigma_k$ provides an algebraic characterization of the motion projected onto the invariant subspace (the genuine eigenmanifold) spanned by a pair of conjugate eigenvectors. For clarity, in what follows, 'eigenmanifold vector' always refers to this real vector $\sigma_k$, being an algebraic concept for the geometric subspace.} theory hypothesis \cite{WY2020,wang2022,WY2022}, which posits that:

\begin{center}
\fbox{%
  \parbox{0.850\textwidth}{%
The dynamic structure of the game's steady-state
system is characterized by the linear superposition of eigenmanifolds, which depends specifically on the eigenvector structure at the Nash equilibrium and is ultimately governed by the game dynamics equations.
  }%
} 
\end{center}

 This theoretical hypothesis, by introducing the construction of eigencycles, enables both theoretical calculation and experimental measurement   simultaneously. In discrete-strategy discrete-time games, it has been repeatedly supported across 17 different settings of real human game experiments during the past 40 years, as shown in the entries of the "Manifold theory validation" column in the bottom block of the Table (\ref{table:existLit}).\\

  However, whether this hypothesis holds in continuous-strategy games remains an open question. The task of this paper is to {verify whether the theoretical hypothesis of game dynamics (eigenmanifold) structure holds, using experimental data from human continuous-strategy games.} That is, to provide a test report for the up block of the Table (\ref{table:existLit}).\\

  For the convenience of readers, the core academic terms used in this paper, along with their mathematical symbols, definitions, and sources, are summarized in section~\ref{sec:eigen_keyword}. 

\subsection{Data\label{sec:data}}

  The research object of this paper is a continuous strategy (location game, or price game) experiment containing six parameter settings, as shown in top rows of Table \ref{table:existLit}, involving both discrete time and continuous time. The data can be obtained from the experimental literature \cite{dan2021price,dan2026Cyclical} publication website.

\begin{table}[!ht]
\centering
\caption{Eigenmanifold in various game protocol \label{table:existLit}}
\begin{tabular}{@{}llp{2.5cm}llllc@{}}
\toprule
Strategy & Time & \centering Game type & No. of & No. of & Experiment & Manifold & Index of \\
space & setting & & stra. & para. & reference & theory  & result fig. \\
&  & &  &  &  &  validation &  \\
\midrule 
\multirow{3}{*}{Contin.} & Discrete & Single-population & $\infty$ & 
$2$ & CFH2021 \cite{dan2021price} & This paper & 
\ref{fig:sub:dan2021_4},\ref{fig:sub:dan2021_6},\ref{fig:sub:dan2021_8}\\
 &  & Single-population & $\infty$ & * & CF2003\cite{dan2005}
 & This paper&
 \ref{fig:dan2003_T_v} \\

 & Contin. & Single-population & $\infty$ & $4$ & CFGS2026
 \cite{dan2026Cyclical} & This paper &
 \ref{fig:sub:Dan2026_bin4},\ref{fig:sub:Dan2026_bin6},\ref{fig:sub:Dan2026_bin8} \\
\midrule

\multirow{7}{*}{Discrete} & \multirow{7}{*}{Discrete} & Two-population fixed pairing & $8$ & $1$ & O1987\cite{ONeill1987} 
& Q2021\cite{WY2020} & \ref{fig:sub:O1987}\\

& & Single-population & $4$ & $1$ & Z2021A4\cite{2021Shujie} 
&  W2023\cite{Wang2022A4}  & \ref{fig:sub:Z2021A4}\\

& & Two-population  random match & $6$ & $1$ & B2001G6\cite{Binmore2001Minimax} & 
 S2024 & \ref{fig:sub:B2001G6}\\

& & Single-population & $5$ & $1$ & Q2021Y5\cite{2021Qinmei}
&  Q2021Y5\cite{2021Qinmei} &\ref{fig:sub:y5_8} \\

& & Two-population random match  & $8$ & $1$ & B2001G7\cite{Binmore2001Minimax}& 
 WY20\cite{WY2020} &\ref{fig:sub:oneill_b_8} \\

& & Single-population   & $5$ & $5$ & W2023\cite{WY2022} 
& W23\cite{WY2022} & \ref{fig:A5}\\

& & Two-population fixed pairing & $8$ & $4$ & Ok2013\cite{Yoshitaka2013Minimax} & 
 Q2021\cite{2021Qinmei} &\ref{fig:sub:oneill_iitt_8} \\

& & Two-population fixed pairing & $16$ & $3$ & WW2023\cite{WW2023pulse} & 
 WW2023\cite{WW2023pulse}&\ref{fig:eigencycle_spectra}  \\
\bottomrule
\end{tabular}
\end{table}

% \subsection{方法\label{sec:Approach}}
\subsection{Approach\label{sec:Approach}}

% 验证动力学(流形)结构理论假说的核心方法仍是动力系统的本征系统分析
% (即流形分析或谱分析)。 为使得该方法可
% 普遍适用于连续策略空间博弈, 我们首先需要将空间离散化, 然后通过线性化
% 和本征系统方法得到本征流形。 这样下一步的理论假说的验证就成为可
% 能. 为验证结果的鲁棒性, 可以采取不同精度格点的离散方法以考察可靠
% 性.本研究过程简述如下(详见附录：方法)。

 The core method for verifying the hypothesis of dynamic (manifold) structure theory remains the eigen-system analysis of dynamical systems (i.e., manifold analysis or spectral analysis). To make this method generally applicable to games with continuous strategy spaces, we first need to discretize the space, and then obtain the eigenmanifolds through linearization and eigen-system methods. This makes the subsequent verification of the theoretical hypothesis possible. To verify the robustness of the results, discretization methods with different grid precisions can be adopted to examine reliability. The research process is briefly reported as follows (for details, see Appendix: Methods for details).

% \subsubsection{步骤 0：选取博弈的实验参数\label{sec:step0}}
\subsubsection{Step 0: Select Existed Experimental Game as Sample  \label{sec:step0}}

% 本文用到的参数是$6$组参数, 两种来自 CFH21 \cite{dan2021price}，
% , 四组来自 CFGS \cite{dan2026Cyclical}。
% 每次取定一组博弈参数作为本征流形理论假说验证的起点。

 The treatments used in this paper are $6$ sets of parameters, two from CFH2021 \cite{dan2021price} (2 discrete time treatments), and four from CFGS2026 \cite{dan2026Cyclical} (4 continuous time treatment). Each time, one set of game experiment data is taken as the starting point for verifying the eigenmanifold theory hypothesis.

%\subsubsection{步骤 1：离散化\label{sec:step1}}
\subsubsection{Step 1: Discretization\label{sec:step1}}

% 离散化是常用的方法是将位置区间 $[x_{\min}, x_{\max}]$离散化为
% $N$ 个离散点 $\{x_0,x_1, x_2, \dots, x_N\}$作为玩家的
% \textbf{纯策略}。此时状态空间为 $N$ 维单纯形，每个社会状态向量
% $p = (p_1,\dots,p_N)$ 表示选择各策略的玩家比例。为研究系统的
% 鲁棒性以及为与已有的离散系统的结果比较, 本文主要报告
% $N=\{4,6,8, 10, 20,..., 100\}$ 的情形. 注意, 以下两点

  Discretization is a common method that discretizes the position interval $[x_{\min}, x_{\max}]$ into $N$ discrete points $\{x_0,x_1, x_2, \dots, x_N\}$ as the players' \textbf{pure strategies}. In this case, the state space is an $N$-dimensional simplex, and each social state vector $p = (p_1,\dots,p_N)$ represents the proportion of players choosing each strategy. To study the robustness of the system and to compare with the results of existing discrete systems, this paper mainly reports the cases of $N=\{4,6,8, 10, 20,..., 100\}$. Note the following two points:

% \begin{itemize}
%     \item  在实验测量中，将实验中选择了低于下界$x_{\text{min}}$的
%     被试数据样本归入纯策略$x_1$, 将高于上界$x_{\text{max}}$的纳
%     入纯策略$x_N$. 
%     \item  将系统转换为 $N$ 个格点的方案, 在概念上可借助有限
%     格点振动系统的振动与模态关系的直观图像. 有限格点振动系统的任
%     意振动可分解为各阶模态(固有振型)的线性叠加。每个模态对应一
%     个固有频率和特定的振动形态，系统的实际运动由各模态按一定权重
%     线性组合而成，模态之间相互独立。
%     %
%     % 如果一个模态相对衰减慢(即本征值实部$\Re (\lambda_k)$)相对较大)，
%     % 或者一个模态频率高(即本征值虚部的绝对值$|\Im (\lambda_k)|$)相对较大)，
%     % 则该模态的权重较大，成为相对主导项。 这是本文理论验证
%     % 的判据的数学基础. 
% \end{itemize}

\begin{itemize}
    \item In experimental measurements, data samples from subjects that fall below the lower bound $x_{\text{min}}$ are classified into the pure strategy $x_1$, while those above the upper bound $x_{\text{max}}$ are classified into the pure strategy $x_N$.
    \item This is a scheme for converting the continuous strategy space into $N$ grid (discrete set with Euclidean position), which conceptually can be aided by the intuitive image of the relationship between vibrations and modes in a finite lattice vibration system. Any vibration of a finite lattice vibration system can be decomposed into a linear superposition of various modes (natural vibration shapes). Each mode corresponds to a natural frequency and a specific vibration pattern. The actual response of the system is formed by a linear combination of the modes with certain weights, and the modes are independent of each other.
    %
    % If a mode decays slowly (i.e., the real part of the eigenvalue $\Re (\lambda_k)$) is relatively large), or if a mode has a high frequency (i.e., the absolute value of the imaginary part of the eigenvalue ($|\Im (\lambda_k)|$) is relatively large), then the weight of that mode is larger, making it a relatively dominant term. This is the mathematical basis for the criterion verified theoretically in this paper.
\end{itemize}

\begin{figure}[!ht]
    \centering
    \includegraphics[width=0.8\linewidth]{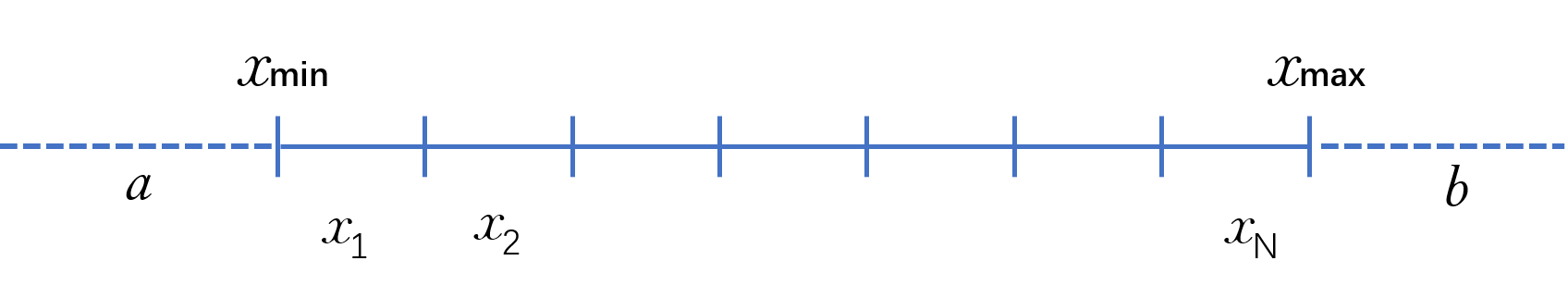}
    \caption{Discretization method}
    \label{fig:discrete_appr}
\end{figure}

% \subsubsection{步骤 2：本征流形向量\label{sec:step2}}
\subsubsection{Step 2: Construct Manifold Vector\label{sec:step2}}

% \paragraph{计算离散情形下纳什均衡}
% 对于离散化，可以将连续系统的在$N$个离散小区间上的概率密度函数
% ($pdf$)积分, 作为均衡策略概率$p_i^*$, 要求满足$\sum p_i^*= 1$.
% 在本文涉及的实验, 原文\cite{dan2021price,dan2026Cyclical}都提供有清晰的理论纳什均衡的连续函数形式, 而对于不同$N$离散方案, 保持了分布形态和均衡收益不变\footnote{在replicator dynamics中, 均衡分布对应着特征值为0的特征向量.}。 
% %

 \paragraph{Computing Nash Equilibrium in the Discrete Set}
For discretization, the probability density function (pdf) of the continuous system over the $N$ discrete small intervals can be integrated to obtain the equilibrium strategy probabilities $p_i^*$, which must satisfy $\sum p_i^* = 1$. In the experiments involved in this paper, the original references \cite{dan2021price,dan2026Cyclical} provide clear continuous functional forms of the theoretical Nash equilibrium, and for different $N$ discretization schemes, the equilibrium strategy distribution and  payoffs are preserved\footnote{In replicator dynamics, the equilibrium distribution corresponds to the eigenvector with eigenvalue zero.}.

% \paragraph{计算本征模态系统(本征值, 本征向量, 本征圈和流形流形向量)}
% 采用复制子动力学
% %
% \footnote{在Sandholm\cite{2011Sandholm}
% 提及的其它常见的动力学模型, 根据已有的在不同的博弈采用不同的动力学模型的验证
% \cite{Wang2022A4,2021Qinmei,2021Shujie}, 在特征流形结构上没有显著差异, 作为展示对特征流形的理论验证, 这里只讨论这个动力学模型}, 
% %
% 在均衡点处的雅
% 可比矩阵$J$为 $N\times N$ 矩阵,计算$J$的本征值$\lambda_k$
% 和本征向量 $\xi_k$. 然后, 计算理论本征圈$\sigma^{mn}$ 条件范围为~ 
% $(N-1 \geq m>n>1)$ . 这给出了所有在 $(m,n) $二维子空间上预测的理
% 论本征流形$\sigma_k$向量 (方法详见附录第
% \ref{sec:eigenmanifold}节), 作为本文理论假说
% 验证方程 (\ref{eq:main}) 的理论值基础。
% %

 \paragraph{Computing the eigen mode system (eigenvalues, eigenvectors, eigencycle and theoretical manifold vectors)}  
Using the replicator dynamics  
\footnote{In Sandholm \cite{2011Sandholm}, other common dynamic models are mentioned. Referring to previous studies \cite{dan2005,wang2017,Wang2022A4,2021Qinmei,2021Shujie}, in which the eigenmanifold structure has been verified under different dynamic models and in different games (CF2003 \cite{dan2005}; the cyclic motion of discrete-strategy games \cite{wang2017}; the discrete-strategy games A4 \cite{Wang2022A4}, Qimei \cite{2021Qinmei} and Shujie \cite{2021Shujie}), no significant difference in the structure of eigenmanifolds has been found. As a demonstration of the theoretical validation of eigenmanifolds, only the replicator dynamics is discussed here.},
the Jacobian matrix $J$ at the equilibrium point is an $N\times N$ matrix. Compute the eigenvalues $\lambda_k$ and eigenvectors $\xi_k$ of $J$. Then, compute the theoretical eigencycle $\sigma^{mn}_k$ condition range ~  
$( 1 \leq m<n \leq N)$. Referring to the term 'Order of Manifold Vector
Component' in Table \ref{tab:eigen_keyword}, this yields all theoretical eigenmanifold vectors $\sigma_k$ (see Appendix Section \ref{sec:eigenmanifold} for details), serving as the theoretical basis for verifying the theoretical hypothesis Eq. (\ref{eq:main}).

% \paragraph{从实验数据中测量实验流向量}
% 在每个实验时间步上，首先将每个被试在连续策略空间上的策略映射到
% 离散化后的纯策略， 而得到离散化策略分布 $p_i(t)$。然后在每个时间
%  步上, 对每个二维子空间$(m,n)$，计算全部时间段上的实验平均子空间角动量
%  $\bar{L}^{mn}$。按照定义, 依顺序排列$\bar{L}^{mn}$ 构
%  成实验流形向量$\bar{L}$. 方法详见附录第\ref{sec:exp_L}节。

 \paragraph{Measuring the experimental manifold vector from experimental data}
At each experimental time step, the strategy of each subject in the continuous strategy space is first mapped to the discretized pure strategy, yielding the discretized strategy distribution $p_i(t)$. Then, at each time step, for each two-dimensional subspace $(m,n)$, the  average experimental manifold vector $\bar{L}^{mn}$ over the entire time period is calculated. By referring to the term 'Order of Manifold Vector
Components' in Table \ref{tab:eigen_keyword}, the sequentially arranged $\bar{L}^{mn}$ form the experimental manifold vector $\bar{L}$. For details, see Appendix Section \ref{sec:exp_L}.

% \subsubsection{步骤 3：统计检验方法\label{sec:step3}}
\subsubsection{Step 3: Statistical Test Method\label{sec:step3}}

% \paragraph{统计检验模型设定}
\paragraph{Statistical Test Model Specification}
%%%%%  <need translate \begin>
% 对于每个实验系统与离散化参数 \(n\)，建立如下多元线性回归模型：
%%%%%  <need translate \end>  

 For each experimental system and discretization parameter,  $N$, the following multiple linear regression model (F-test) is applied:
\begin{equation} 
\label{eq:main}
    \bar{L} = \sum_{k=1}^N c_k \sigma_k + \varepsilon,
\end{equation}

% 其中 \(\sigma_k\) 为第 \(k\) 个本征流形向量(计算见公式~(\ref{eq:mn_order}))，\(\bar{L}\) 为实验流向量(公式~(\ref{eq:L_order}))，\(c_k\) 为参与因子，\(\varepsilon\) 为误差项。

  where \(\sigma_k\) is the \(k\)-th eigenmanifold vector (see equation~(\ref{eq:mn_order})), \(\bar{L}\) is the experimental manifold vector (see equation~(\ref{eq:L_order})), \(c_k\) is the participation factor, and \(\varepsilon\) is the error term.  
The hypothesis of Eq.~(\ref{eq:main}) is stated for the manifold vectors, whereas the eigen decomposition of the linearized dynamics is stated for the motion of the state vector (see Appendix \ref{Prerequisites}, Eq.~(\ref{eq:eigdeco2}); the equivalent of these two objects when cyclic motion are test \cite{WY2020,2021Qinmei}. Detail explanation see Appendix \ref{pra:equilvant2obj}.

\begin{enumerate}
    \item \textbf{Full model test} Full model significance test
Null hypothesis \(H_0: c_1 = c_2 = \cdots = c_N = 0\), using \(F\) test to calculate the \(p\)-value. If \(p < 0.05\), reject the null hypothesis, concluding that the set of eigenmanifolds as a whole has explanatory power for the experimental data. Plot a scatter plot of the experimental values \(\bar{L}\) versus the model fitted values \(\sum c_k \sigma_k\) to visually demonstrate the fitting performance of the model under different experimental systems and discretization methods.

% \item \textbf{Principal characteristic manifold existence test} \(\sigma_{1st}\) is the characteristic manifold corresponding to the eigenvalue with the largest imaginary part.

\item \textbf{Principal eigenmanifold Existence Test} Define $\sigma_{\text{max}}$ as the eigenmanifold corresponding to the eigenvalue with the largest imaginary part (i.e., the highest oscillation frequency and longest strategy space wave length); This is \textbf{one wave going round} mode in strategy space, why this mode is the dominant one is explained in Appendix~\ref{sec:app_dom}. If this manifold dominates the system dynamics, its regression coefficient $c_{\text{max}}$ should be significantly different from zero , and the corresponding $t$-statistic $t(c_{\text{max}})$ is reported in Appendix~\ref{app:c1c2c3}.
% , and removing $\sigma_{\text{max}}$ from the full model should lead to a significant decline in the model's explanatory power. 

% \item   \textbf{Marginal contribution of the principal/subprincipal eigenmanifold}
% Let \(\sigma_{\text{max}}\) and \(\sigma_{2nd}\) correspond to the eigenmanifolds of the eigenvalues with the largest and second largest imaginary parts, respectively. To test their independent contributions:
%  Construct a reduced model: remove \(\sigma_{\text{max}}\) (or \(\sigma_{2nd}\)) from the full model;
%     Compare the \(F\) test \(p\)-values of the two models. If the \(p\)-value increases significantly after removal (e.g., \(\Delta p > 0.01\) and relative increase \(>100\%\)), then the manifold is considered to have a statistically significant marginal contribution.

    \item     \textbf{Eigenmanifold weights order}~
    % The weight for difference mode  is ordering by their related imaginary part of eigenvalue $|\operatorname{Im}\lambda_k|$  (see Appendix \ref{sec:app_dom} for explanation), that is, in Eq. (\ref{eq:main}), $c_{max} > c_{2nd} > c_{3rd}$ should hold.   \\
    The weights of the different modes are ordered by the imaginary part of the eigenvalue to which they belong, $\lvert\operatorname{Im}\lambda_k\rvert$ (see Appendix~\ref{sec:app_dom} for the mechanism). That is, in Eq.~(\ref{eq:main}), $c_{\text{max}}>c_{\text{2nd}}>c_{\text{3rd}}>0$, each coefficient being significant (Eq.~\ref{eq:criterion_c123}); the evidence is the coefficient tables of Appendix~\ref{app:c1c2c3}.
\end{enumerate}

% \subsection{总结\label{sec:Summary}}
\subsection{Summary\label{sec:Summary}}

% 连续策略博弈, 和离散策略博弈一样\cite{Wang2022A4,WY2020,WY2022,2021Qinmei,2021Shujie}, 其博弈动力
% 学结构依然是由本征流形线性叠加而成.

 Continuous strategy games, like discrete strategy games\cite{Wang2022A4,WY2020,WY2022,2021Qinmei,2021Shujie}, still have their game dynamics structure formed by the linear superposition of eigenmanifolds.

% 基于纳什均衡的系统动力学方法所提供的、由本征流形表示的动力学结构，
% 可以预言真实系统演化行为的重要本征。这一点在连续策略博弈上与离散策略博弈一致。

 The dynamical structure provided by the Nash equilibrium-based system dynamics method and represented by eigenmanifolds can predict important features of the evolutionary behavior of real systems. This is consistent across both continuous-strategy and discrete-strategy games.

% 尽管真实的人类博弈中演化过程是复杂的, 但并非混乱无序，纳什均衡
% 概念仍然是博弈系统的灵魂, 而本征流形理论假说可能提供了对博弈运动的
% 洞察.

 Although the evolutionary process in real human games is complex, it is not chaotic or disordered. The concept of Nash equilibrium remains the soul of the game system, and the eigenmanifold theory hypothesis may provide insight.

%
%
% \section{结论\label{sec:results}}
\section{Results\label{sec:results}}
%%%%%  <need translate \begin> 
%  本节提供主要结论及其证据,  
% 同时展示本征流形概念为连续策略博弈提供的新知识。
%%%%%  <need translate \end> 

This section presents the main results and the supports,   and also demonstrates the insights provided by the  manifold concept for continuous-strategy games.
% \begin{itemize}
% % \item  结果汇总见表(\ref{tab:result_sum}) . 
% % 它归纳了三项核心结果：流形假说的验证、主本征流形的存在性及其边际贡献、
% % 次本征流形的存在性及其边际贡献。
% % 每一项结论均附有结论说明、与已有结果的对比、对应的可视化图表位置以及统计检验的引用节或数据表列。
% \item  结果图见(图\ref{fig:eCy_2026},\ref{fig:eCy_2021}), 解释如下:  --\textcolor{red}{这里图和表的编号做了ref,确认一下}
% \begin{itemize}
%     \item 对连续策略, 离散时间情形, 对于不同的间隔$N=4,6,8$
%     的结果图见 \ref{fig:sub:dan2021_4},\ref{fig:sub:dan2021_6},\ref{fig:sub:dan2021_8}. 二个不同的参数(Unstable, Stable)的
%     情形下, 实验值在本征流形空间的线性展开的F-检验值, 见表  \ref{tab:F-test_p1} 
%     \item 对连续策略连续时间情形, 对于不同的间隔$N=4,6,8$的结
%     果图见 \ref{fig:sub:Dan2026_bin4}, \ref{fig:sub:Dan2026_bin6}, \ref{fig:sub:Dan2026_bin8}. 四个不同的参数($\gamma=2, 3.6,4.4,
%     6$的情形下, 实验值在本征流形空间的线性展开的F-检验值,见表 \ref{tab:F-test_p1} 
% \end{itemize}
%  由于对连续策略空间的离散化, 是把博弈映射成 $N$个离散策略, 这样
%  本征流形假说的呈现就与离散博弈是一致的. 所以, 我们提供已有的相应
%  的离散策略的 本征流形假说理论检验的图. 其中, $N$ =4, 6, 8,   
%  对应于4,6,8离散策略且离散时间博弈的理论实验, 数据来自论文 \cite{Wang2022A4,WY2020}
%  (这里6策略博弈采用的数据是来自Binmore的实验, 二人群三策略的设置; 
%  理论部分来自浙江大学<<演化博弈理论与实验(2023-2026)>>教程材料). 
% \end{itemize}

\subsection{Main results}\label{sec:sum_result}
\begin{itemize}
\item The main results are summarized in Table (\ref{tab:result_sum}). It summarizes three core results: the verification of the manifold hypothesis, the existence of the Principal eigenmanifold and the order of eigenmanifold weight.
% and its marginal contribution, and the existence of the secondary eigenmanifold and its marginal contribution. 
Each conclusion is accompanied by a description of the conclusion, a comparison with existing results, the location of the corresponding visualization chart, and the reference section or data table column for statistical tests.
\item The result figures are shown in (Figures \ref{fig:eCy_2026},~\ref{fig:eCy_2021}), with explanations as follows:
\begin{itemize}
    \item For the continuous strategy, discrete-time case, with different intervals $N~=~4,~6,~8$, the result figures are shown in Figure ~\ref{fig:sub:dan2021_4},~\ref{fig:sub:dan2021_6},~\ref{fig:sub:dan2021_8}. Under two different parameter settings (Unstable, Stable), the F-test values for the linear expansion of experimental values in the eigenmanifold space are shown in Table \ref{tab:F-test_full} .
    \item For the continuous strategy, continuous-time case, with different intervals $N=4,~6,~8$, the result figures are shown in  \ref{fig:sub:Dan2026_bin4}, \ref{fig:sub:Dan2026_bin6}, \ref{fig:sub:Dan2026_bin8}. Under four different parameter settings ($\gamma~=~2,~3.6,~4.4,~6$), the F-test values for the linear expansion of experimental values in the eigenmanifold space are shown in Table \ref{tab:F-test_full} .
\end{itemize}
Due to the discretization of the continuous strategy space, the game is mapped into $N$ discrete strategies, so the presentation of the eigenmanifold hypothesis should be comparable  with those of discrete games with same strategy numbers. Therefore, we provide existing corresponding figures for the test of the eigenmanifold hypothesis for discrete strategies. Here, $N=4,~6,~8$ correspond to experiments with 4, 6, and 8 discrete strategies. The data are from the papers \cite{Wang2022A4,WY2020} (the data for the 6-strategy game are from Binmore's experiment \cite{Binmore2001Minimax}, with a two-population, three-strategy setting, which the theoretical part is from the teaching materials of the course "Evolutionary Game Theory and Experiments (2023-2026)" at Zhejiang University).
\end{itemize}

\begin{table}[!ht]
\centering
\caption{Summary of Main Results}\label{tab:result_sum}
\begin{tabular}{|c|m{2.4cm}|m{5.5cm}|m{2.2cm}|m{2.5cm}|}
\hline
No. & Result & Explanation & Visualization & Statistical Test \\
\hline
1 &
Manifold hypothesis supported &
The dynamic structure of the game is a linear combination of eigenmanifolds in significant &
Figures 2a, 2b, 2c, 3a, 3b, 3c &
Section \ref{sec:sta_full} \newline \& Table (\ref{tab:F-test_full})   \\
\hline
2 &
Existence of principal \newline eigenmanifold &
The principal eigenmanifold exists significantly, in the sense of the largest participation factor shown in Section~\ref{sec:sta_c123}
&
- &
Section \ref{sec:sta_max} \newline \& Table (\ref{tab:F-test_1st}) 
%only $\sigma_{\text{max}}$  % and Table (\ref{tab:F-test_M1st}) w/o-1st manifold 
\\
\hline
% 3 &
% Existence of secondary \newline eigenmanifold &
% The secondary eigenmanifold marginal contribution  is significant &
% - &
% Section \ref{sec:sta_2nd} \newline \& Table (\ref{tab:F-test_M2nd}) w/o-$\sigma_{\text{2nd}}$ submanifold \\
% \hline
3 &
Ordering of the \newline eigenmanifold weights &
The participation factors of the three leading eigenmanifolds satisfy  $c_{\text{max}}>c_{\text{2nd}}>c_{\text{3rd}}>0$ for $N\ge50$ in every treatment, and the ratios satisfy $c_{\text{2nd}}/c_{\text{max}}\subset[0.37,0.67]$ and $c_{\text{3rd}}/c_{\text{max}}\subset[0.10,0.44]$ &
- &
Section \ref{sec:sta_c123} \newline \& Tables in Appendix (\ref{app:c1c2c3})    \\
\hline
\end{tabular}
\end{table}

% \subsection{统计检验结果说明\label{sec:statis_Test_result}}
\subsection{Interpretation of Statistical Test Results\label{sec:statis_Test_result}}
% \subsubsection{1. 理论本征流形向量与实验流形向量线性关系的稳健
% 性\label{sec:sta_full}}
\subsubsection{Linear Dependence between Theoretical and Experimental Manifold\label{sec:sta_full}}

% \textbf{样本说明：} 统计样本来源见表\ref{table:existLit}中的"连续策略"的6个实验参数设置.  每个实验参数设置(\texttt{treatment})对应一
% 个独立系统，将连续位置空间
% 离散化为 $N$个策略$(N=10,20,\dots,100)$, 构成一个检验结果。每个检验方法一样, 回归模型的自变量为理论预测的本征流形模型\texttt{eigencycle}值 $\sigma_k,~k\in (1....N)$(包含$N$个向量, 每个向量有$C_N^2$个标量元素)，
% 因变量为实验测量的平均实验流向量 $\bar{L}$(包含$C_N^2$个元素的向量)。

 \textbf{Sample description:} The statistical sample sources are shown in Table \ref{table:existLit} for the six experimental parameter settings of the "continuous strategy". Each experimental parameter setting (\texttt{treatment}) corresponds to an independent system, where the continuous position space is discretized into $N$ strategies $(N=10,20,\dots,100)$, indexed as \texttt{bin\_num} in Table \ref{tab:F-test_full}, forming one F-test result, respectively.  The independent variable of the regression model is the theoretical manifold set $\sigma_k,~k\in (1....N)$ (containing $N$ vectors, each vector having $C_N^2$ scalar elements), and the dependent variable is the experimentally measured average experimental manifold vector $\bar{L}$ (a vector containing $C_N^2$ elements).

% \textbf{判据：} 线性叠加假说成立的前提是理论向量与实验向量之间存
% 在统计上显著的线性关系，即全模型回归的 $p$ 值应小于 0.05，且随着
% 离散化精度增加($k$ 增大)$p$ 值应持续下降，表明关系更加稳健。
 \textbf{Criterion:} The premise for the validity of the linear superposition hypothesis is that there exists a statistically significant linear relationship between the theoretical vector set and the experimental vector, i.e.,
 % the $p$-value of the full model  regression, F-test, should be less than 0.05, and as the discretization precision increases (bin\_num  increases), the $p$-value  in Table \ref{tab:F-test_full}) should continue to decrease, indicating a more robust relationship.
 the $p$-value of the full-model $F$ test, reported as \texttt{pValue\_full} in Table~\ref{tab:F-test_full}, should be less than $0.05$ in every $(\texttt{treatment},\texttt{bin\_num})$ combination. 
 % … so the $F$ statistic — and with it the $p$-value — need not change monotonically with $N$ (in Table~\ref{tab:F-test_full}

%%%%%  <need translate \begin> 
% \textbf{统计检验方法} 采用普通最小二乘(OLS)线性回归，并使用$F$检验评估
% 本征流形模型的整体显著性。
% \textbf{$p$ 值：} 表中 \texttt{pValue\_full} 即为该$F$ 检验
% 对应的 $p$ 值。

%%%%%  <need translate \end> 

\textbf{Statistical test method:} $F$-test, the ordinary Least Squares (OLS) linear regression,  is employed to evaluate the overall significance of the eigenmanifold model.
\textbf{$p$-value:} In the table, \texttt{pValue\_full} is the $p$-value corresponding to this $F$-test.

% \textbf{结果表明：} 随着 $k$ 从 10 增至 100，所有 
% \texttt{treatment} 下的 \texttt{pValue\_full} 均从
% $10^{-6} $量级降至 $10^{-80} $乃至 0(例如 
% \texttt{treatment=6, bin\_num=100} 时 $p=0$)。在所有 
% $60$ 个($6$ \texttt{treatment} × $10$ 
% \texttt{bin\_num})组合中，最大 $p$ 值 $<0.01$，无一例外。
% 这为动力学结构是本征流形的线性叠加假说的稳健性提供了支持。

 \textbf{Result:} As  $N$, as (bin\_num) in Table \ref{tab:F-test_full} increases from 10 to 100, the \texttt{pValue\_full} under all 6  \texttt{treatment} conditions decreases from the order of $10^{-6}$ to $10^{-80}$ or even 0 (for example, when \texttt{treatment=6, bin\_num=100}, $p=0$). Among all $60$ combinations ($6$ \texttt{treatment} $times$ $10$ \texttt{bin\_num}), the maximum $p$ value is $<0.01$, without exception. This provides support for the robustness of the hypothesis that the dynamical structure is a linear superposition of eigenmanifolds.

% \subsubsection{2. 主本征流形的存在性与主导作用
% \label{sec:sta_max}}
\subsubsection{Existence of the Principal manifold\label{sec:sta_max}}

% \textbf{样本说明：} 样本同第1点。在每个
% $(\texttt{treatment}, \texttt{bin\_num}) $组合中，全模型
% 包含所有理论本征流形向量(对应不同本征值)。主本征流形定义为具有
% 最大复数本征值虚部(即最高周期频率)的本征向量所对应的流形。

 \textbf{Sample description:} The sample is the same as above, by $(\texttt{treatment}, \texttt{bin\_num})$ combination. The Principal eigenmanifold is defined as the manifold corresponding to the eigenvector with the largest imaginary part of the complex eigenvalue (i.e., the highest periodic frequency). Appendix~\ref{sec:app_dom} shows that this mode is the dynamically dominant one, so the statistical test to this mode is on target. 
 % One verification for the $\sigma_{\text{max}}$ manifold is conducted. 
 	The statistical test below therefore bears on the dynamically dominant mode, not on an arbitrarily chosen one.

% \begin{enumerate}
% \item 主本征流形存在证据显著；---\textcolor{red}{图和表的ref，确认一下}

%     \textbf{判据：} 若主本征流形存在，则全模型回归应统计显著，即
%     回归 $F$ 检验的 $p$ 值满足$p_{\text{1st}} < 0.05$在所有
%     参数实验和所有离散方法中均应成立。
    
%     \textbf{统计检验方法：} 采用 $F$ 检验判断全模型(包含所有理论
%     本征流形向量)的显著性。表(\ref{tab:F-test_p1})中 \texttt{pValue\_1st} 即为单
%     模型的 $p$ 值。
    
%     \textbf{$p$ 值：} 在所有 60 个组合中，
%     $\texttt{pValue\_full} < 0.05$ 严格成立，无一例外。见表(\ref{tab:F-test_p1})中的1st 表. 
    
%     \textbf{结果表明：} 主本征流形在所有实验条件下均表现出极强的统计
%     显著性，为其存在性提供了定量证据。 

% \item 主本征流形对实验解释的边际贡献显著;
    
%     \textbf{判据：} 若主本征流形存在且为主导，则从全模型中
%     \textbf{单一剔除}该流形后，模型的解释力应下降，具体表现为
%     回归 \(p\) 值增大，即  
%     %
%     \begin{equation}
%      p_{\text{full}} - p_{\text{missing\_1st}} < 0   
%     \end{equation}
%     %
%     在所有组合中均应成立。

%     \textbf{统计检验方法：} 比较全模型与剔除主本征流形后的嵌套模型
%     的 $F$ 检验 $p$值。表(\ref{tab:F-test_p2})中 \\
%     \texttt{pValue\_single\_missing\_max}即为剔除后模型的
%     $p$值。
    
%     \textbf{$p$ 值：} 在所有 $60$ 个组合中，
%     $\texttt{pValue\_1st} - 
%     \texttt{pValue\_single\_m1st} < 0 $严格成立，
%     无一例外  
    
%     \textbf{结果表明：} 这定量证明了主本征流形对实验数据具有重要的
%     边际贡献。
% \end{enumerate}

% \begin{enumerate}
% \item
The existence of the Principal characteristic (1st) manifold is statistically significant;

    \textbf{Criterion:} If the Principal  manifold exists, {its regression on the experimental manifold vector} should be statistically significant, i.e., the $p$-value of the regression $F$-test satisfies $p_{\text{1st}} < 0.05$ for all parameter experiments and all discretization methods.  The $F$-test is used to verify the significance of  the Principal  manifold  model.  

    \textbf{Result:} The Principal eigenmanifold exhibits strong statistical significance under all F-test, providing quantitative evidence for its existence. (In all 60 samples, $\texttt{pValue\_1st} < 0.05$ holds strictly, without exception. See the Table (\ref{tab:F-test_1st})).

% \item The marginal contribution of the Principal eigenmanifold   is significant;

%     \textbf{Criterion:} If the Principal eigenmanifold exists and is dominant, then \textbf{removing only} this manifold from the full model should reduce the model's explanatory power, specifically manifested as an increase in the regression $p$-value, i.e.,
%     %
%     \begin{equation}\label{eq:criterion_m1st}
%      p_{\text{full}} - p_{\text{missing\_1st}} < 0   
%     \end{equation}
%     %
%     should hold for all combinations. The $F$-test $p$-values of the full model (see Table \ref{tab:F-test_full}) and the $F$-test $p$-value with only the $\sigma_{\text{max}}$  manifold  is removed from full model (see Table(\ref{tab:F-test_M1st})) is compared.
 
%     \textbf{Result:} Results quantitatively demonstrates that the principal eigenmanifold has a significant marginal contribution to the experimental data. In all 60 samples, the criterion Eq. \ref{eq:criterion_m1st} is satisfied, without exception.
 % \end{enumerate}

  \textbf{Explanation:} This test is necessary because, in these experiments, as Appendix~\ref{sec:app_dom} shows, their cumulative-utility operator is a spatial low-pass filter: its gain and frequency both peak at the longest wavelength, making principal-manifold dominance natural and expected.

\subsubsection{Eigenmanifold Weights ordering \label{sec:sta_c123}}

% \textbf{样本说明：} 同第 1 点。$c_k$ 为式(\ref{eq:main}) 中的参与因子，
% 即实验流形向量在第 $k$ 个本征流形向量上的回归系数。
% 待办：① 附录里加上 \label{app:c1c2c3}；② 与 item 3 的“无例外”措辞统一（本段的 Result 含例外）。

\textbf{Sample description:} Same as in Section~\ref{sec:sta_full}. Referring to Eq. \ref{eq:main}, the participation factor $c_k$ is the regression coefficient of the experimental manifold vector on the $k$-th eigenmanifold vector in Eq.~(\ref{eq:main}).

\textbf{Criterion:} If the dynamics is a superposition of eigenmanifolds dominated by the long-wave modes, then the participation factors of the three leading eigenmanifolds---those of the largest, the second largest and the third largest imaginary parts, i.e.\ of one, two and three waves around the strategy space---should be positive and decrease in magnitude,
\begin{equation}\label{eq:criterion_c123}
c_{\text{max}} > c_{\text{2nd}} > c_{\text{3rd}} > 0 .
\end{equation}
The criterion is stated on the ratios of the coefficients rather than on their absolute values: both the scale and the number of eigenmanifold vectors depend on the discretization $N$, so the coefficients themselves are not comparable across $N$, whereas their ratios are.

\textbf{Result:} Eq.~(\ref{eq:criterion_c123}) holds in $54$ of the $60$ experimental combinations (six treatments $\times$ ten discretization levels). Details of the numeric result are shown in  Appendix~\ref{app:c1c2c3}. 
\begin{itemize}
    \item All six exceptions lie at the coarse end of the discretization range: for $\gamma=2$ the third participation factor turns negative at $N=10$ and $N=20$; and for $\gamma=4.4$ the second factor exceeds the first at $N=10$ to $N=40$. 
    \item For $N \geq 50$ the ordering holds in all $36$ remaining combinations, and the ratios there are narrowly bounded: at $N=100$,  
    $$c_{\text{2nd}}/c_{\text{max}} \subset [0.37,0.67] ~~ \text{comparable with $c_{\text{2nd}}/c_{\text{max}}$ = 1/2 in the model}$$ and $$c_{\text{3rd}}/c_{\text{max}} \subset  [0.10,0.44] ~~ \text{comparable with $c_{\text{3rd}}/c_{\text{max}}$ = 1/3 in the model}$$ across the six treatments (for the model $c_k/c_1 = 1/k$ result, Eq. \ref{eq:1_k}, see Appendix  \ref{sec:app_dom}).\\ {Note: $c_{1},c_{2},c_{3}$ in the tables of Appendix~\ref{app:c1c2c3} are $c_{\text{max}},c_{\text{2nd}},c_{\text{3rd}}$ here, ordered by $|\operatorname{Im}\lambda_k|$.}
    \item Individually, $c_{\text{max}}$ is significant in all $60$ combinations, $c_{\text{2nd}}$ in $58$ (the two exceptions are again at $N=10$). 
The coefficient tables, the significance of every coefficient, and the ill-conditioning diagnostic are given in Appendix~\ref{app:c1c2c3}.
\end{itemize}

\textbf{Explanation:} The exceptions are not distributed at random: they all lie at the coarse end of the discretization range, where the gravest mode of the cumulative force field---the wave whose dominance is derived in Appendix~\ref{sec:app_dom}---cannot be resolved (at $N=10$ only three or four conjugate pairs exist, against $41$ to $49$ at $N=100$), so the leading modes are not separated and their weights may cross or change sign; the criterion is asymptotic, and it is the large-$N$ combinations that test it. The measured ratios are not only ordered but of the predicted size, $c_{\text{2nd}}/c_{\text{max}}$ bracketing $1/2$ and $c_{\text{3rd}}/c_{\text{max}}$ bracketing $1/3$, the residual spread being expected because the $1/k$ law rests on mode-independent damping, whereas the experimental games may have mode-dependent damping. \\
% Finally, since the design matrix is ill-conditioned at large $N$, only the leading coefficients are interpretable and only they are reported; this does not affect the tests of items 1--3, which concern the fitted values $\sum_k c_k\sigma_k$ rather than individual coefficients.\\

%%
%% 20260914-1321 end
%%

 In brief, the above statistical analysis provides  empirical support for the hypothesis that "the dynamic structure of games can be linearly superimposed by eigenmanifold," and clarifies the quantitative role of the leading (dominant) eigenmanifold in explaining human experimental game data.

\subsection{Advantages of the Eigenmanifold Method\label{sec:Comparative_Advantages}}

% 本征流形理论假说的关键是本征圈, 它同时是本征流形的理论计算对象, 
% 又是实验测量的对象. 实验流形向量的矩阵表示可以直观展示博弈演化过
% 程中策略间的净转移模式，可以表示为反对称矩阵 $\sigma^{mn}$。其中，行
% $i$($i=1,\dots,8$)表示时刻 $t$的策略, 列 $j$ 表示时刻 
% $t+\Delta t$ 的策略；元素 $\sigma_{ij}$ 为从策略 $i$ 到策略$j$ 的
% 净流量相对值(由本征圈理论值或实验测量值计算得到)。这里$i,j$的值
% $1,2,...,8$表示价格从低到高的策略区间; 若 $\sigma_{ij}>0$，则净流
% 动方向为 $i \to j$；若 $\sigma_{ij}<0$，则净流动方向为 $j \to 
% i$。矩阵满足反对称性 $\sigma_{ji}=-\sigma_{ij}$，对角元$ \sigma_{ii}=0$。在实验的每个矩阵元素$Tmn$是可依据测量方程
% (\ref{eq:experiment_sigma_mn_L})测量的.

 The key of the eigen manifold theory hypothesis is the eigen cycle, which is both the theoretical calculation object of the  eigen  manifold and the object of experimental manifold $\bar{L}$ measurement. \\
 
 The matrix representation of the experimental manifold vector can intuitively display the net transfer patterns among strategies during the game evolution process, and can be expressed as an antisymmetric matrix $\sigma^{mn}$. Here, row $i$ ($i=1,\dots,8$) represents the strategy at time $t$, and column $j$ represents the strategy at time $t+\Delta t$; the element $\sigma_{ij}$ is the relative net flow from strategy $i$ to strategy $j$ (calculated from the theoretical value of the eigencycle or the experimental measurement value in 2d subspace $(i,j)$). Here, the values $1,2,...,8$ of $i,j$ represent the strategy intervals from low to high prices; if $\sigma_{ij}>0$, the net flow direction is $i \to j$; if $\sigma_{ij}<0$, the net flow direction is $j \to i$. The matrix satisfies antisymmetry $\sigma_{ji}=-\sigma_{ij}$, and the diagonal elements $\sigma_{ii}=0$. In the experiment, each matrix element $T^{mn}$, exactly equals to $\bar{L}^{ij}$, can be measured according to the measurement Eq (\ref{eq:experiment_sigma_mn_L}).\\

% 以CFGS的实验{\cite{dan2026Cyclical}}为例,图
% ~\ref{fig:heatmap_g2_g6_exp} 展示了参数 $\gamma=2$ 与
%  $\gamma=6$ 下的净流量热图. 
% 这里
% 说明本征流形理论假说框架的优势. 
% 说明如下:

\begin{figure}[!ht]
\centering
    \includegraphics[width=0.45\textwidth]
    {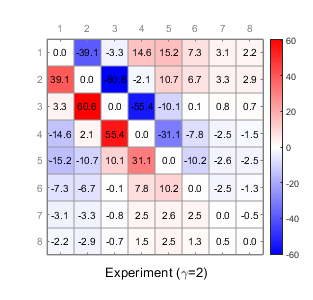} ~~~~~~
    \includegraphics[width=0.45\textwidth]
    {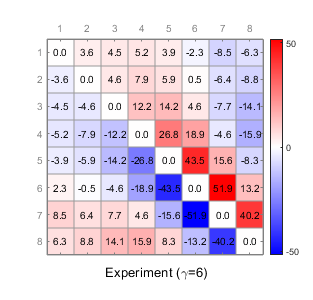}\\
    % 实验值
    % \includegraphics[width=1\linewidth]
    %{heatmap_g2_g6_theo.png}\\
    % 理论值
\caption{
% 实验流形值矩阵表示的热图。左：
% $\sigma_{\text{$\gamma=2$}}$($\gamma=2$)；右：
% $\sigma_{\text{$\gamma=6$}}$($\gamma=6$)。红色(正值)
% 表示净流量从行策略流向列策略，蓝色(负值)表示相反方向。 
Heatmap representation of the experimental manifold value matrix ( in CFGS2026 \cite{dan2026Cyclical}). Left: Experimental manifold $\bar{L}$ ($\gamma=2$); Right: Experimental manifold $\bar{L}$ ($\gamma=6$). Red (positive values) indicates net flow from the row strategy to the column strategy, while blue (negative values) indicates the opposite direction. The experimental sample consists of 3439 rounds observation for $\gamma=2$ and 3445 rounds observation for $\gamma=6$.}
%该样本包含 $\gamma=2$ 的观测数据 3,439 条，$\gamma=6$ 的观测数据 3,445 条。}}
\label{fig:heatmap_g2_g6_exp}
\end{figure}

\begin{figure}[!ht]
\centering
    \includegraphics[width=.45\textwidth]
    {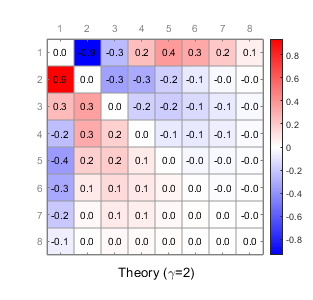} ~~~~~~
    \includegraphics[width=.45\textwidth]
    {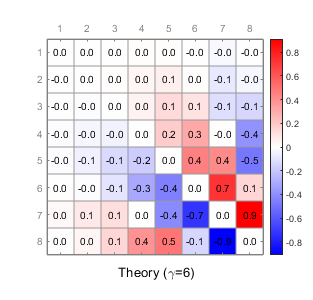}\\ 
\caption{
% 主本征流形对应的理论值矩阵表示的热图。左：
% $\sigma_{max}$($\gamma=2$)；右：
% $\sigma_{max}$($\gamma=6$)。红色(正值)表
% 示净流量从行策略流向列策略，蓝色(负值)表示相反方向。
Heatmap representation of the theoretical matrix corresponding to the Principal eigenmanifold (CFGS2026 \cite{dan2026Cyclical}). Left: $\sigma_{\max}$ ($\gamma=2$); Right: $\sigma_{\max}$ ($\gamma=6$). Red (positive values) indicates net flow from the row strategy to the column strategy, while blue (negative values) indicates the opposite direction. The value for each element $T^{mn}$ is the value   of the \textbf{normalized} theoretical principal manifold $\sigma_{\text{max}}^{mn}$. }
\label{fig:heatmap_g2_g6_max_theo}
\end{figure}

% \subsubsection{实验测量的流形应用\label{sec:Comparative_Advantages_exp measure}}
\subsubsection{Experiment Results Presentation with Eigenmanifold \label{sec:nanifold_for__exp_measure}}

Figure~\ref{fig:heatmap_g2_g6_exp}, being the experiment results of CFGS2026 \cite{dan2026Cyclical},  shows the net flow heatmap under parameters $\gamma=2$ and $\gamma=6$. This illustrates the advantages of the eigenmanifold theory hypothesis framework. The explanation is as follows:

%%%%%  <need translate \begin> 
% 利用本征流形框架得到的结果, 相比于原论文展示的结果，本征流形框架进
% 一步实现了对运动结构的微观化和具体化，具体体现在：
%%%%%  <need translate \end> 

 Compared with the results presented in the original report (CFGS2026 \cite{dan2026Cyclical}), the eigenmanifold framework further achieves a more microscopic and concrete representation of the motion structure, specifically reflected in:

% \begin{enumerate} 
% \item {价格循环的主导方向相反} 对比两个热图可先观察价格近邻策略
% 的$\bar{L}_{mn}$值, 可以看到 

% \begin{enumerate}
%     \item 在 $\bar{L}_{\text{$\gamma=2$}}$ 中，$\sigma_{i,i-1}$ 为正(红
%     色)， 例如 $\bar{L}_{21}=0.4$，$\bar{L}_{32}=0.6$ 
%     (蓝色)。这意味着, 系统整体呈现大量的小幅向下(近邻的)低价位
%     策略的迁移(如 $3\to 2 \to  1$)，伴随着少量的价格向上大幅
%     跃迁(如$1\to4$)。
%     \item 在 $\bar{L}_{\text{$\gamma=6$}}$ 中，$\bar{L}_{i,i+1}$ 为正(红
%     色)，例如 $\bar{L}_{56}$，$\bar{L}_{67}$, $\bar{L}_{78}$。这意味着系统整体呈现大量的小幅向上(近邻的)高价位策略的迁移(如 $6\to7\to8$ 和 $\to4\to 5 \to 6$), 伴随少量从高价向中间
%     价位的大幅跃迁(如$8\to 4$)。
% \end{enumerate} 

%  \item  {价格循环的的活跃区域相反}  两个矩阵的较大绝
%  对值元素所集中的策略区域也显著不同：
 
% \begin{enumerate}
%     \item 在 $\bar{L}_{\text{$\gamma=2$}}$ 中，强度较大的净流量
%     (深色部分)主要位于低间价
%     位策略区域(区间 $1–5$)。这说明其策略向低价位迁
%     移主要发生在低价位策略空间。
%     \item 在 $\bar{L}_{\text{$\gamma=6$}}$ 中，强度较大的净流量
%     (深色部分)
%     主要位于高价位策略区域(区间 $5–8$)。这说明其策略向
%     上迁移主要发生在高价策略空间。
% \end{enumerate} 
% \end{enumerate}

\begin{enumerate} 
\item {The dominant direction of the price cycle is \textbf{opposite}} Comparing the two heatmaps, first observe the $\bar{L}^{mn}$ values of the price nearest-neighbor strategy. It can be seen that 

\begin{enumerate}
    \item In $\bar{L}_{\text{$\gamma=2$}}$, $\sigma_{i,i-1}$ is positive (red), for example $\bar{L}^{21}=0.4$, $\bar{L}^{32}=0.6$ (blue). This means that the system as a whole exhibits a large number of small downward (nearest-neighbor) migrations to low-price strategies (such as $3\to 2 \to 1$), accompanied by a small number of large upward price jumps (such as $1\to4$).
    \item In $\bar{L}_{\text{$\gamma=6$}}$, $\bar{L}^{i,i+1}$ is positive (red), for example $\bar{L}^{56}$, $\bar{L}^{67}$, $\bar{L}^{78}$. This means that the system as a whole exhibits a large number of small upward (nearest-neighbor) migrations to high-price strategies (such as $6\to7\to8$ and $\to4\to 5 \to 6$), accompanied by a small number of large jumps from high to middle prices (such as $8\to 4$).
\end{enumerate} 

 \item  {The active regions of the price cycle are \textbf{opposite}} The strategy regions where the larger absolute values of the two matrices are concentrated are also significantly different:
 
\begin{enumerate}
    \item In $\bar{L}_{\text{$\gamma=2$}}$, the stronger net flows (dark parts) are mainly located in the low-price strategy region (interval $1–5$). This indicates that the migration of strategies to low prices mainly occurs in the low-price strategy space.
    \item In $\bar{L}_{\text{$\gamma=6$}}$, the stronger net flows (dark parts) are mainly located in the high-price strategy region (interval $5–8$). This indicates that the upward migration of strategies mainly occurs in the high-price strategy space.
\end{enumerate} 
\end{enumerate}

% \subsubsection{理论预测的流形应用\label{sec:nanifold_for__the_predit}} 
\subsubsection{Theoretical Prediction Presentation with Eigenmanifold \label{sec:nanifold_for__the_predit} }

% 理论上，对于每个本征向量$k$ ,每一个净流量矩阵元素都存在明确的理论预
% 言值，即本征圈 $\sigma^{mn}_{k}$。根据通过式~
% (\ref{eq:theo_sigma_mn})  计算给定$k$的本征向量的每个二维子空
% 间上的理论循环强度. 图 \ref{fig:heatmap_g2_g6_max_theo} 是根
% 据主导本征向量(对应的本征值的复部最大的本征向量), 可以看到, 在不考
% 虑其它$k$的影响, 单一的主导流形已经预言了$\sigma^{mn}$出给定参数下的运动特
% 征 --- 转动方向和发生区域等.

 Theoretically, for each eigenvector $k$, every element of the net flow matrix has a clear theoretical predicted value, namely the eigencycle $\sigma^{mn}_{k}$. The theoretical cycle intensity on each two-dimensional subspace of the eigenvector for a given $k$ is calculated using equation~(\ref{eq:theo_sigma_mn}). \\

Figure \ref{fig:heatmap_g2_g6_max_theo} is based on the principal  eigenvector (the eigenvector with the largest imaginary part of the corresponding eigenvalue; the justification for using this mode alone is given in Appendix~\ref{sec:app_dom}). It can be seen that, without considering the influence of other $k$ mode, the  1st manifold has already predicted the motion characteristics of $\sigma^{mn}$ under given parameters — such as (1) the direction of rotation, from high price to low slowly and jump back to high quickly in $\gamma = 2$ treatment,  and the region of occurrence is in the low price zone in $\gamma = 2$ treatment, and the reversal in $\gamma = 6$ treatment (in CFGS2026 \cite{dan2026Cyclical}).

% 该段落定义了速度场为概率净流矢量，并提出基于主本征流形(取最大虚部特征值)的理论预测公式 \(v(m)=\sum (n-m)\sigma^{mn}\)，无需数据拟合。通过与CFGP实验测量值对比，发现两者在速度场随策略位置的符号模式(正负交替)及对参数γ的响应趋势上一致，验证了理论的有效性。

%
%%%
\subsubsection{Velocity as Projection of Eigenmanifold \label{sec:nanifold_for_velocity} }
% 速度场是本征流形的投影. 速度指的是经过一个给定态的有向转移的位移平均值, 是社会态空间上表示概率净流的矢量, 它的数学形式可见\cite{wang2017,2011coyness}. 在连续策略空间上, 用速度向量表示不同(价格)策略区段的演化大小和方向, 采用主导流形矢量计算, 图(\ref{fig:velocity_gamma_2_6}a) 是速度场的理论的预言, 它没有经过任何经验数据拟合,也不考虑其它特征流形的权重. 一个本征流形向量$\sigma_k$, 其分量$\sigma_k^{m,n}$, 它对$m$态贡献的速度向量为 $(n-m)\sigma_k^{m,n}$, 假设只考虑主特征流形, 也就是只取最大复部的特征值所对应的特征流形, 最速度场进行理论估算, 那么$m$ 态的速度为
% \begin{equation}\label{eq:velocity}
%     v(m) = \sum_{n=1}^N (x_n-x_m) \sigma^{mn} 
% \end{equation}
% 其中, $x_i$ 是纯策略$i$ 在策略空间上的位置, 也就是价格. 对 CFGS\cite{dan2026Cyclical} 的 $\gamma=2,6$的主本征流形的速度场投影呈现, 取bin\_num = 的结果, 见图(\ref{fig:velocity_gamma_2_6}a) .而
% 图(\ref{fig:velocity_gamma_2_6}b) 是对应的实验测量值, 其测量方法可见\cite{wang2017,2011coyness}. 在这里可表示为
% \begin{equation}\label{eq:velocity_exp}
%     v_{\text{exp}}(m) = \sum_{n=1}^N (x_n-x_m) \bar{L}^{mn} 
% \end{equation}
% 理论与实验的一致性体现在, 对于相对变化模式(峰谷位置、正负号分布、跨参数的反转特征), 理论预言与实验测量结果定性匹配。这也支持了特征流形理论假设的有效性。
%%% 
%

\begin{figure}
    \centering
    \includegraphics[width=0.6\linewidth]{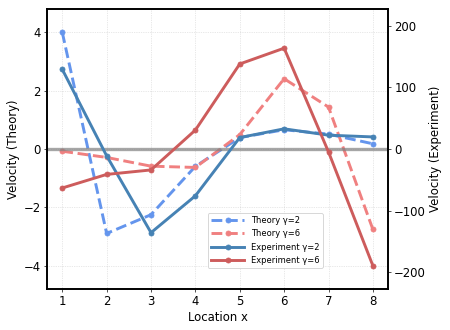}
    \caption{
    % 不同价格区位内的价格变化速度(包括大小和方向)(a)理论图,虚线表示, 根据公式(\ref{eq:velocity}), 采用主导流形矢量计算出; (b)实验测量值, 根据公式(\ref{eq:velocity_exp}). 
    Velocity of price changes across different price locations (including magnitude and direction). (a) Theoretical plot (dashed lines), computed using the dominant manifold vectors according to Eq.~(\ref{eq:velocity}); (b) Experimental measurements (solid line) according to Eq.~(\ref{eq:velocity_exp}).}
    \label{fig:velocity_gamma_2_6}

% % velocity(:,[1 4])

% ans = Theo gamma 2 | 6
% 
%     3.9995   -0.0764
%    -2.8941   -0.2888
%    -2.2519   -0.5882
%    -0.5941   -0.6320
%     0.3913    0.4937
%     0.6604    2.4086
%     0.5095    1.4383
%     0.1794   -2.7551
 
% velocity(:,[1 4])
% 
% ans = = Exp gamma 2 | 6
% 
%   129.3327  -63.2830
%   -11.0585  -41.2356
%  -135.7890  -33.9534
%   -75.9710   30.7904
%    18.6018  138.0733
%    32.7876  163.7369
%    22.4703   -4.6934
%    19.6261 -189.4351
\end{figure}

\begin{figure}[!ht]
    \centering
    \begin{subfigure}[b]{0.42\linewidth}
        \centering
        \includegraphics[width=\linewidth]
        {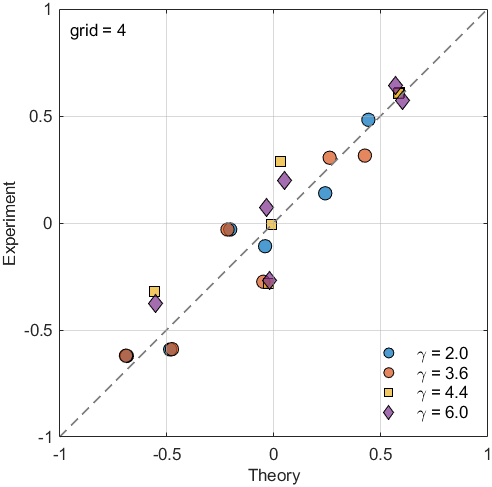}
        \caption{CFGS2026\_bin4}
        \label{fig:sub:Dan2026_bin4}
    \end{subfigure}
    \hfill
    \begin{subfigure}[b]{0.42\linewidth}
        \centering
        \includegraphics[width=\linewidth]{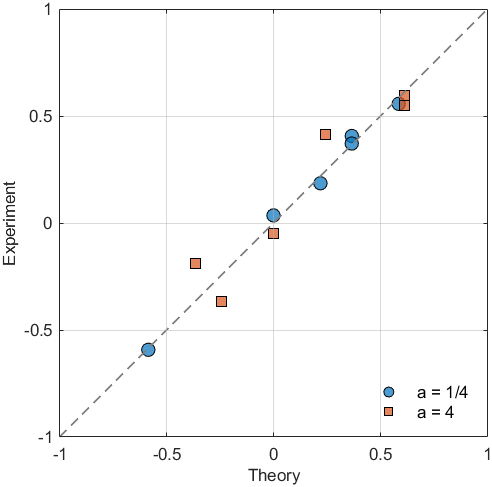}
        \caption{Z2021A4}
        \label{fig:sub:Z2021A4}
    \end{subfigure}
    \\
    \begin{subfigure}[b]{0.42\linewidth}
        \centering
        \includegraphics[width=\linewidth]
        {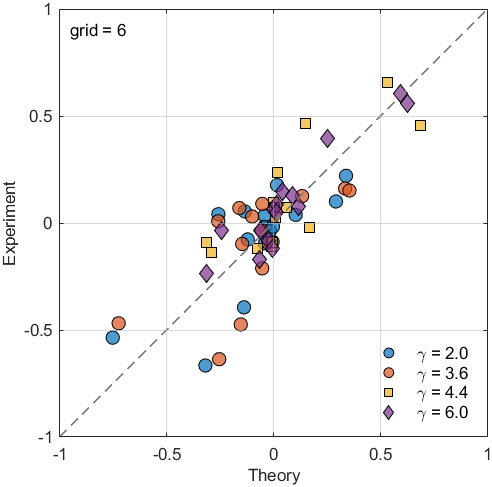}
        \caption{CFGS2026\_bin6}
        \label{fig:sub:Dan2026_bin6}
    \end{subfigure}
    \hfill
    \begin{subfigure}[b]{0.42\linewidth}
        \centering
        \includegraphics[width=\linewidth]
        {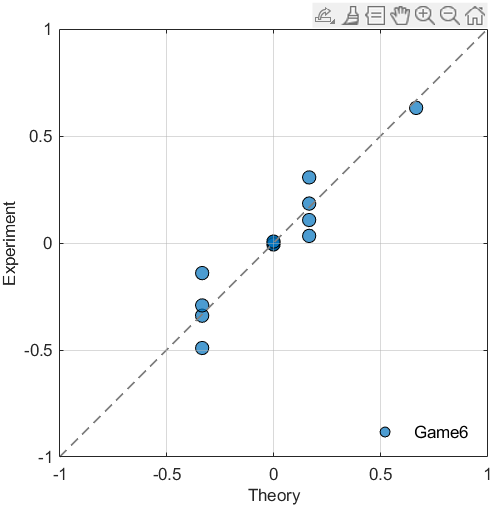}
        \caption{B2001G6}
        \label{fig:sub:B2001G6}
    \end{subfigure}
    \\
    \begin{subfigure}[b]{0.42\linewidth}
        \centering
        \includegraphics[width=\linewidth]
        {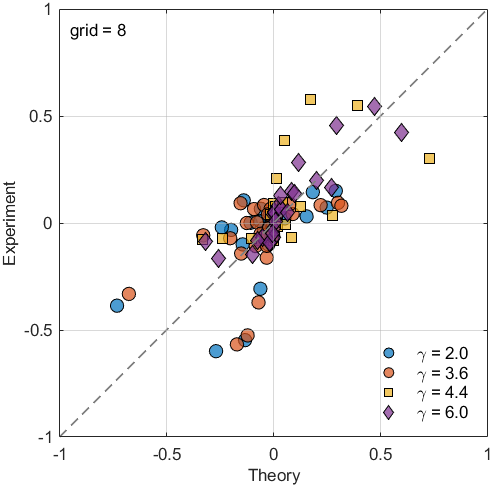}
        \caption{CFGS2026\_bin8}
        \label{fig:sub:Dan2026_bin8}
    \end{subfigure}
    \hfill
    \begin{subfigure}[b]{0.42\linewidth}
        \centering
        \includegraphics[width=\linewidth]
        {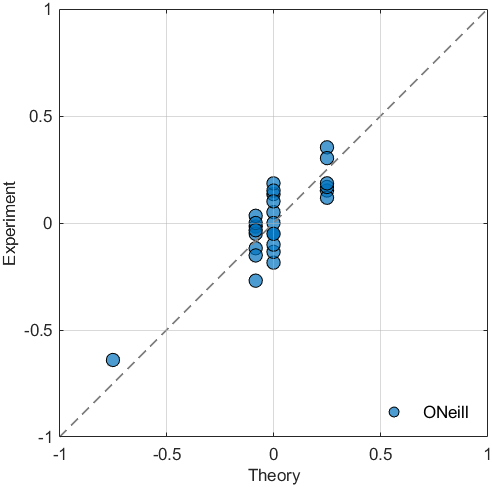}
        \caption{O1987}
        \label{fig:sub:O1987}
    \end{subfigure}
\caption{Comparison of theoretical and experimental eigencycles: The left column subfigures ((a), (c), (e)) show the fitting results of the discretized (CFGS2026\cite{dan2026Cyclical}) experiment with bin\_num 4, 6, and 8; the right column subfigures ((b), (d), (f)) show the fitting results of existing experiments (Z2021A4\cite{2021Shujie}, B2001G6\cite{Binmore2001Minimax}, and O1987\cite{ONeill1987}, respectively). The subfigure labels correspond to the experimental references listed in Table~\ref{table:existLit}.}
\label{fig:eCy_2026}
\end{figure}

\begin{figure}[!ht]
    \centering
    
    \begin{subfigure}[b]{0.42\linewidth}
        \centering
        \includegraphics[width=\linewidth]
        {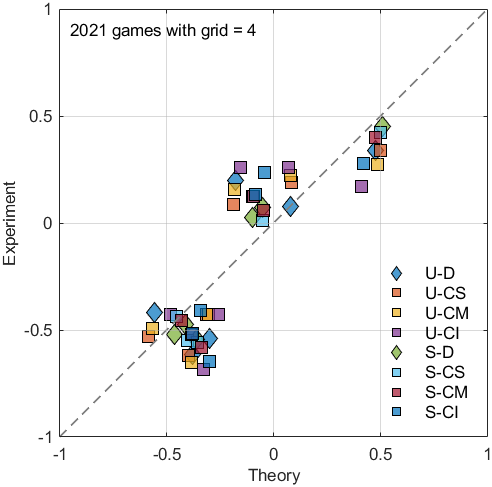}
        \caption{CFH2021\_bin4}
        \label{fig:sub:dan2021_4}
    \end{subfigure}
    \hfill
    \begin{subfigure}[b]{0.42\linewidth}
        \centering
        \includegraphics[width=\linewidth]
        {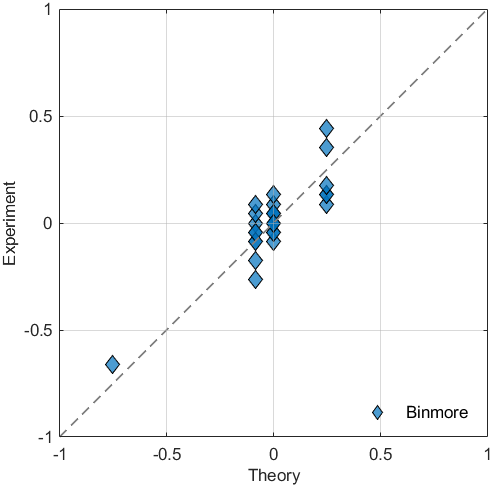}
        \caption{B2001G7}
        \label{fig:sub:oneill_b_8}
    \end{subfigure}
    \\
    
    \begin{subfigure}[b]{0.42\linewidth}
        \centering
        \includegraphics[width=\linewidth]
        {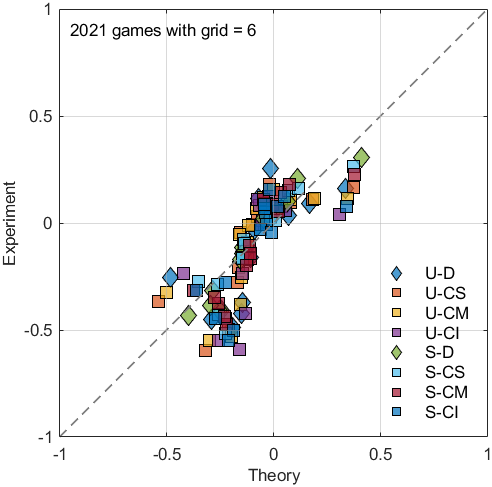}
        \caption{CFH2021\_bin6}
        \label{fig:sub:dan2021_6}
    \end{subfigure}
    \hfill
    \begin{subfigure}[b]{0.42\linewidth}
        \centering
        \includegraphics[width=\linewidth]
        {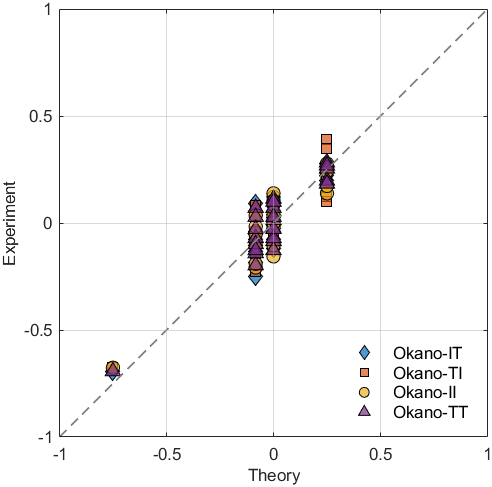}
        \caption{Okans2013}
        \label{fig:sub:oneill_iitt_8}
    \end{subfigure}
    \\
    
    \begin{subfigure}[b]{0.42\linewidth}
        \centering
        \includegraphics[width=\linewidth]
        {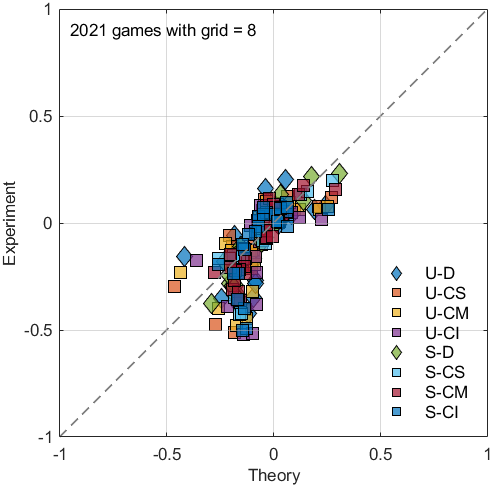}
        \caption{CFH2021\_bin8}
        \label{fig:sub:dan2021_8}
    \end{subfigure}
    \hfill
    \begin{subfigure}[b]{0.42\linewidth}
        \centering
        \includegraphics[width=\linewidth]{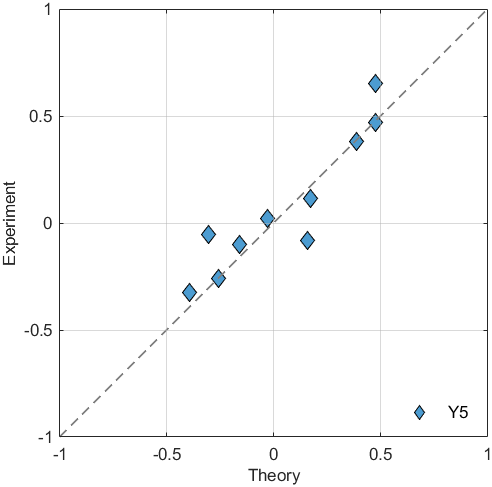}
        \caption{W2023}
        \label{fig:sub:y5_8}
    \end{subfigure}
    
\caption{Comparison of theoretical and experimental manifold: The left column subfigures ((a), (c), (e)) show the fitting results of the discretized (CFH2021\cite{dan2021price}) experiment with bin\_num 4, 6, and 8; the right column subfigures ((b), (d), (f)) show the fitting results of existing experiments (B2001G7\cite{Binmore2001Minimax}, Okans2013\cite{Yoshitaka2013Minimax}, and Q2021Y5\cite{2021Qinmei}, respectively). The subfigure labels correspond to the experimental references listed in Table~\ref{table:existLit}.}
\label{fig:eCy_2021}
\end{figure}

The velocity field is the projection of the eigenmanifold on state space. Velocity refers to the average displacement of directed transitions passing through a given state, and it is a vector in the social state space representing the net probability flow. Its mathematical formulation can be found in \cite{wang2017,2011coyness}. In a continuous strategy space, the velocity vector is used to represent the magnitude and direction of evolution across different (price) strategy intervals. For example, using the principal manifold vector, dash lines in Figure (\ref{fig:velocity_gamma_2_6}) shows the theoretical prediction of the velocity field, which is obtained without any empirical data fitting and without considering the weights of other eigenmanifolds. The method is explained follow.

\paragraph{One-dimensional velocity-field representation of the  manifold.}
For an eigenmanifold vector $\sigma_k$'s  components $\sigma_k^{m,n}$, its contribution to the velocity of state $m$ is $(n-m)\sigma_k^{m,n}$.  As justified in Appendix~\ref{sec:app_dom}, assuming only the Principal eigenmanifold (i.e., taking only the eigenmanifold corresponding to the eigenvalue with the largest imaginary part) is used to theoretically estimate the velocity field, then the velocity of state $m$ is given by

\begin{equation}\label{eq:velocity}
    v_{\text{theo}}(m) = \sum_{n=1}^N (x_n-x_m) \sigma^{mn}.
\end{equation}

where $x_i$ is the position of pure strategy $i$ in the strategy space, which corresponds to prices in CFGS~\cite{dan2026Cyclical} and CFH~\cite{dan2021price}. For the CFGS \cite{dan2026Cyclical} experiments with $\gamma=2,~6$, the velocity field projections of the Principal eigenmanifold are presented (with a given bin\_num = 8) in Figure (\ref{fig:velocity_gamma_2_6}) in dash lines. Figure \ref{fig:velocity_gamma_2_6} shows the corresponding experimental measurements, whose measurement method can be found in \cite{wang2017,2011coyness}. In this study case, we have 
\begin{equation}\label{eq:velocity_exp}
    v_{\text{exp}}(m) = \sum_{n=1}^N (x_n-x_m) \bar{L}^{mn} 
\end{equation}
Figure \ref{fig:velocity_gamma_2_6}, 
the consistency between theory and experiment is reflected in the relative patterns of change (peak and valley positions, sign distribution, and reversal characteristics across parameters), where the theoretical predictions qualitatively match the experimental measurements. This consistency also supports the validity of the eigenmanifold theory hypothesis. \\

In addition, we notice that, the velocity field of the 
post price market (CF2003 \cite{dan2003,dan2005}) can be obtained from their manifold in data. The main results of   the velocity field and the explanation are shown in Fig. (\ref{fig:dan2003_T_v}) and its caption. The advantage of the eigenmanifold presentation is significant, when comparing with the analysis (Table 1 transition matrices in \cite{dan2005}, section 4.3. Hypothesis tests, where significantly different from zero at the marginal
10\% level (Wilcoxon test)). 

\paragraph{Two-dimensional velocity-field representation of the  manifold (Rock-Paper-Scissors representation of manifold).}
If the continuous strategy space is discretized by tertiles(i.e.,
Rock-Paper-Scissors), consequently, the players' strategies, originally continuous
observations, are converted into choices among three strategies (R, P, S), which yields an equivalent
structure of a multi-player Rock-Paper-Scissors game \cite{MIT2014Best,wang2014social}. Then two-dimensional velocity-field representation of the characteristics manifold (Rock-Paper-Scissors representation) can be obtained. The figure \ref{fig:RPS-dan2021-D} shows the experimental velocity fields obtained from the CFH2021 \cite{dan2021price} data for the Stable-Discrete Time treatment and Unstable-Discrete Time Treatment, computed using equation (\ref{eq:velocity_exp}) with the experimental manifold from the treatments, respectively. 

% \subsubsection{讨论\label{sec:discus}}
\section{Discussion\label{sec:discus}}
\subsection{Two Argument on dynamics experiment \label{sec:two_argu}}

Two arguments are discussed here. 

% \paragraph{线性化问题}~
% 基于将连续纳什均衡离散化并在其附近进行线性化的特征流形假说，在真实系统分布长期处于远离纳什均衡的情况下，它是否仍然适用？答案是：在已有的离散策略真人实验中，群体状态常明显偏离纳什均衡(如 Binmore 等，2001)，但基于纳什均衡线性化的本征流形方法保持有效。本文及前期研究共计20个不同参数设置的真人博弈实验(14个离散策略 + 6个连续策略)统一采用角动量测量和速度场, 二者都基于本征流形，未对任何实验单独定制分析方法。结果表明：测量到的实验流形向量 $\bar{L}$ 与理论本征流形向量 $\sigma_k$ 在全模型回归中始终呈现极显著的线性关系。因此，即使系统状态远离纳什均衡，线性叠加假说依然成立。这是经验结果。

\paragraph{Argument on linearization}
 For the eigenmanifold hypothesis,  based on discretizing the space and linearizing around Nash equilibrium, does it remain applicable when the real system distribution stays far from the Nash equilibrium for extended periods? 
 Answer: In existing discrete-strategy human-subject experiments, group states often deviate significantly from the Nash equilibrium (e.g., Binmore et al., 2001), yet the eigenmanifold method based on linearization remains valid. This paper, together with previous studies, covers a total of 20 human-subject game experiments with various settings (17 discrete-strategy + 6 continuous-strategy), all uniformly using experimental angular momentum and velocity field \cite{wang2017}, without customizing the analysis method for any experiment. The results show that the experiment and the theoretic exhibit a highly significant consistency. Therefore, we suggest, even when the system state moves far from the Nash equilibrium, the linear superposition hypothesis still provides framework to capture the motion, e.g., by the principle eigenmanifold. This is an empirical result. 

% \paragraph{东一榔头西一棒子的问题}
%  博弈动力学研究在理论与实验二者之间, 相互约束和验证的关系,长期以来, 没有得到系统性的增强和改善。 
%  %
%  其一，理论方面：已有成果大多局限于各自的假设性的博弈场景，每项研究依赖各自的动力学模型（例如最佳反应动态、自适应学习或参数经过特定选择的仿真），其理论预测往往缺乏跨场景的泛化能力(或称外部有效性)；不同工作独立开展，鲜有对实验数据进行系统性跨实验跨场景比较，缺乏普遍性的对实验的预言能力。 
%  %
%  其二，实验方面：常用的检测方法常常是一种统计指标，其构造依赖于特定假设，而非从动力学方程直接推导, 未能与底层动力学系统建立直接的理论联系，这使得现有分析多为定性判断（例如“振幅更大”“顺时针旋转”），而缺乏可明确计算、具有明确动力学意义的实验观察值系统。

%  通过大范围排查已有结果, 推进理论与实验的内在的统一, 这是这个经历了近40年领域迫切的工作.
\vskip 0.5cm

\textbf{Argument on First a hammer east, then a stick west.} The mutual constraint and validation between theory and experiment, in the field of evolutionary game dynamics, has long lacked systematics.
\begin{enumerate}
    \item First, on the theoretical side: most existing results are confined to their own hypothetical game scenarios, with each study relying on its own dynamic model (e.g., best-response dynamics, adaptive learning, or simulations with specifically chosen parameters). Their theoretical predictions often lack cross-scenario generalization (or external validity). Different works are carried out independently, with few systematic cross-experiment, cross-scenario comparisons of experimental data, and thus there is a lack of general and strong  predictive power regarding experimental observation.
    \item Second, on the experimental side: commonly used measurement methods are often statistical indicators whose construction depends on specific assumptions rather than being directly derived from dynamic equations, or presented in phase diagram strictly. Thus the reports fail to establish a direct theoretical link with the underlying dynamical system. As a result, existing analyses are mostly qualitative judgments (e.g., "larger amplitude", "clockwise rotation"), lacking a system of experimentally observable quantities, which could be sufficient self-consistency and complicity,  that can be clearly computed and carry clear dynamical meaning. Thus experiments can provide little stress to theory, and then little power to push dynamics theory forward. 
\end{enumerate}
% By conducting a broad survey of existing results and advancing the internal unification of theory and experiment, this is an urgent task for a field that has developed over nearly forty years. 
In sum, currently, there is little consensus between theory and experiment, and neither can provide the other with the driving force for progress; this is the status quo. By conducting a broad survey of existing results and advancing the internal unification of theory and experiment, this is an urgent task for a field that has developed over nearly forty years.

\vspace{1em}
\subsection{Related literature}
\paragraph{Relation to spectral economics literature}
\label{sec:disc-spectral}
% \textcolor{red}{ref check}
Eigenvector extensively appears economics  by social network economics, resting on \textbf{real} eigenvector in spectral economics \cite{lizardo2025eigenvectors,golub2026spectral,hidalgo2021economic}. The
Economic Complexity Index is the second eigenvector of a symmetrised proximity
matrix and serves static ranking and clustering
\cite{hidalgo2021economic,caldarelli2012network};
network-learning theory reads the
left eigenvector as the consensus limit
\cite{golubJackson2010naive,golubJackson2012homophily},
key-player and contagion results use the
spectral radius \cite{elliottGolubJackson2014financial};
The two recent surveys \cite{golub2026spectral,hidalgo2021economic} collect precisely
these real-spectrum tools; the only tradition that reads a {complex} eigenvector
is the static matrix data analysis line \cite{ermannShepelyansky2015google} which has provide significant message on economics structure; However, it    
never appear in economic dynamics tool box, nor engagement with observations over time series, nor
does it simultaneously serve the theoretical and the empirical side, see Table \ref{tab:rank-vs-time}.

\begin{table}[ht]
\centering
\small
\caption{The two  eigenvectors }
\label{tab:rank-vs-time}
\begin{tabular}{p{2.0cm} | p{4.5cm} | p{5.5cm}}
\hline
 & Real eigenvector
 & Complex eigenvector   \\
\hline
 Quantity carried & component magnitudes and signs; no phase & component phases {and}
magnitudes; phase differences between components \\
\hline
Measure & Perron vector, Katz--Bonacich,
  ECI, eigenvector centrality, Fiedler vector
 &   eigenmanifold vector, eigencycle scale, winding number, time frequency, space frequency, dominant mode, wave velocity, velocity field \\
\hline
Economic question & \emph{Who matters} ~~static ranking, centrality,
key players,  complexity ranking, network influence, consensus
weights  & \emph{Who moves when} ~~within-cycle schedule, transmission chain, lead--lag, transmission lags, policy timing \\
% \hline
% Economic question &&  \\
\hline
% Time & no, a static eigenvector component ranking& yes, a dynamic eigenvector component ranking\\
Time & only as a long-run limit (stationary distribution, stable age distribution); no
intrinsic frequency & intrinsically: a complex frequency, and through the component phases an order
in time \\
\hline
Ranking & the ranking comes from a {structural snapshot}
& the ranking comes from the {utility driving dynamics}; the order
is joint across all units, carries the global motion. \\
\hline
\end{tabular}
\end{table}

\paragraph{Relation to the time-series literature}
\label{sec:disc-timeseries}
 At its root, eigenmanifold serves as the time-series predictor and measurer.
The complex eigenvectors with components (phase order and winding number) determines their  engagement.
Three contrasts position it to time series tools: where Granger's typical spectral shape
\cite{granger1966typical} is a {temporal  low-pass filter}, ours is its \textbf{spatial low-pass filter}; where the Kolmogorov cascade \cite{kolmogorov1941local} is a nonlinear,
flux-conserving transfer,  our eigencycle spectrum is fixed by the dispersion of a
accumulative linear operator and as the result,  the fastest oscillation sits at the longest wavelength; and where business-cycle
measurement reads complex pairs as an oscillation signature
\cite{chow1968acceleration,chowLevitan1969nature}, we read from them a within-cycle
joint ordering of all variables to identify the full strategy space motion.

\paragraph{Relation to wave mechanics in physics}
\label{sec:disc-physical}

Since the finding of the eigencycle in the O'Neill(1987) game \cite{WY2020,2021Qinmei}, which has been found robustly in various discrete strategy games (see Table~\ref{table:existLit}), we have searched for a closely  related physics system to establish the dynamics picture for the games played by human. By the mathematics, the principal mode is the  complex eigenvector with \textbf{one winding number} (see Appendix \ref{sec:app_dom}).  It seems It seems that a physical counterpart should exist. Unfortunately, we report the absence of a match in our survey.   \\

The ``one wave going round'' as a wave mechanics object is not new \cite{alonsoBarEchebarria2016nonlinear}.
What we did not find in the physical wave systems we surveyed is the combination of
a self-sustained wave on a conserved density, a dispersion $\omega\propto1/k$ in which
the longest mode in full strategy space with the oscillates fastest in local strategy space, and that same mode dominating the flow and
carrying most of the eigenmanifold (high dimensional angular-momentum) variance ($c_1>c2>...$). Physical relatives --- spiral and
reentrant waves, Rossby and drift waves \cite{zaqarashvili2021rossby},
conserved-density granular and traffic waves \cite{helbing2001traffic}, deep-water
gravity and rogue waves \cite{dudley2019rogue}, and non-local continuum neural fields
\cite{bressloff2012spatiotemporal} --- satisfy only part of this list, because
dispersive-wave theory fixes $\omega(k)$ through {local} operators
\cite{elHoefer2016dispersive}; the combination we observe calls for the global (non-local force, but full strategy space long range) 
cumulative force driving by the game economic utility function (or payoff function), disregarding whether the game designer knew or not at its very beginning of an experiment.

\vspace{1em}

\begin{figure}[!ht]
    \centering
    \includegraphics[width=0.45\linewidth]{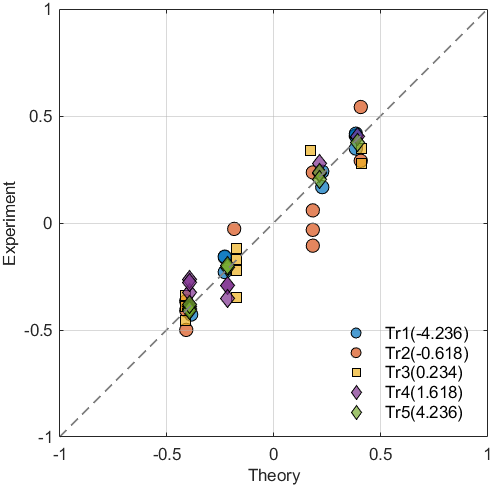}
    \caption{Comparison of theoretical and experimental manifold: W2023\cite{WY2022} experiment. The corresponding experimental reference is listed in Table~\ref{table:existLit}.}
    \label{fig:A5}
\end{figure}

\begin{figure}[!ht]
    \centering
    \adjustimage{width=0.47\linewidth, valign=c}{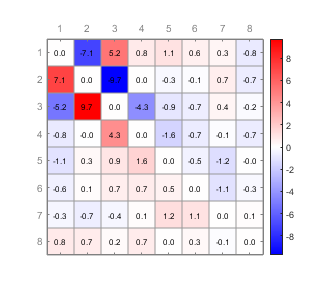}
    % \hfill
    \adjustimage{width=0.42\linewidth, valign=c}{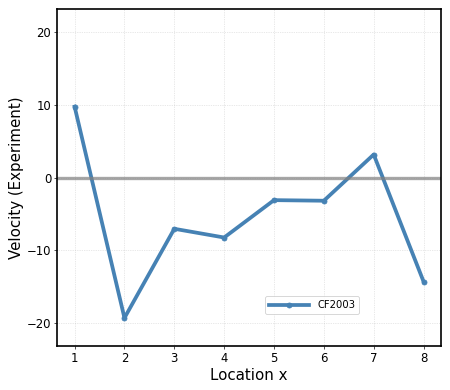}
    \caption{Matrix presentation of the experimental manifold (left) and its velocity vector field projection (right). Data from the Posted-Offer Market (Seller pricing with buyer search) in lab experiment (CF2003 \cite{dan2003,dan2005}). The corresponding experimental reference is listed in Table~\ref{table:existLit}. 
    Here, we pool together all data, which can be obtained (910 rounds observation in 11 data files), without distinguish price search $q=(0,1/3,2/3,1)$ and cost $c=(20, 60)$ parameters. Given that there are considerable records of price convergence to the maximum value of 0 for $q=0$ and to the maximum value of 200 for $q=1$ in the data, the left and right boundaries of the velocity field exhibit a declining pattern, which is understandable from the perspective of the velocity-field definition (for instance, if one round's transaction price is 150 and the next is 200, and the price remains at 200 due to convergence, then a negative velocity is observed at the point 200). Since in this experiment the cases with searcher proportion $q=(1/3,2/3)$ and search cost $\text{cost}=(20,60)$ share similar parameter with the experimental setup of CFH2021 \cite{dan2021price} and also align with the theoretical expectation for $\gamma=2$ in CFGS2026 \cite{dan2026Cyclical}, as shown by the blue dashed line in Figure~\ref{fig:velocity_gamma_2_6}, we omit the detailed theoretical correspondence here.
    % 在这里, 我们将所有的 搜寻者比例 $q=(0,1/3,2/3,1)$的数据累积在一起, 不加以区分;考虑到数据中存在相当比例的 $q=0$的价格收敛到最大值0的记录 和  $q=1$的价格收敛到最大值200的记录, 所以, 速度场的左边边界和右边边界出现下降情况, 从速度场定义的角度这是可以理解的(比如, 叫一轮交易价格是150, 下一轮是200; 由于收敛而停留在到200,那么在100的这一点就观察到负的速度.). 由于在该实验中, 参数 搜寻者比例 $q=(1/3,2/3)$  和搜寻成本$\text{cost}=(20, 60)$的情形特征和 CFH2021 的实验设置相似, 也和 CFGS2026的 $\gamma=2$的理论期望值相似, 如同图(\ref{fig:velocity_gamhma_2_6} )中的蓝色的虚线. 详细的理论对应结果,在这里就不赘述.
    % https://gitee.com/shan-lixia/dan2025/blob/master/Dan2005%E6%95%B0%E6%8D%AE%E5%A4%84%E7%90%86/T-Velocity-data.md
    }
    \label{fig:dan2003_T_v}
\end{figure}

\begin{figure}[!ht]
    \centering
    
    \begin{subfigure}[b]{0.48\linewidth}
        \centering
        \includegraphics[width=\linewidth]
        {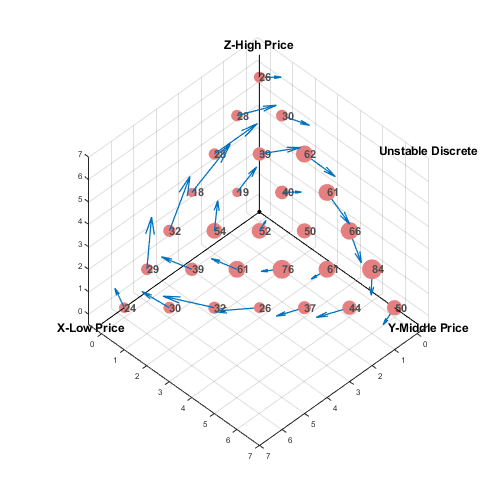}
        \caption{CFH2021\_Discrete\_Unstable}
        \label{fig:sub:Dan2021DUnew}
    \end{subfigure}
    % \hfill
    \begin{subfigure}[b]{0.48\linewidth}
        \centering
        \includegraphics[width=\linewidth]
        {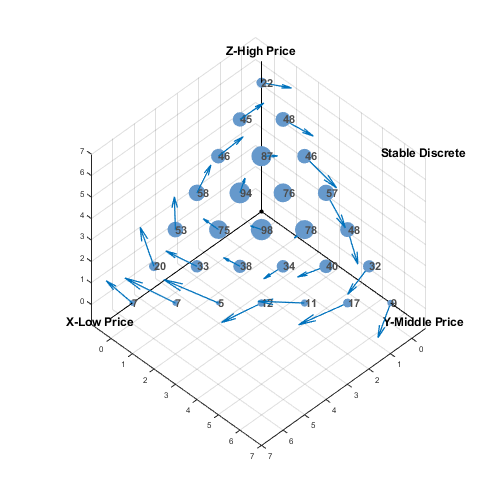}
        \caption{CFH2021\_Discrete\_Stable}
        \label{fig:sub:Dan2021DSnew}
    \end{subfigure}
    % \\
    
    % \begin{subfigure}[b]{0.42\linewidth}
    %     \centering
    %     \includegraphics[width=\linewidth]
    %     {Dan2021DU.png}
    %     \caption{Discrete\_Unstable\_method2}
    %     \label{fig:sub:Dan2021DU}
    % \end{subfigure}
    % \hfill
    % \begin{subfigure}[b]{0.42\linewidth}
    %     \centering
    %     \includegraphics[width=\linewidth]
    %     {Dan2021DS.png}
    %     \caption{Discrete\_Stable\_method2}
    %     \label{fig:sub:Dan2021DS}
    % \end{subfigure}
    \caption{Two-dimensional velocity-field representation of the experimental manifold (Rock-Paper-Scissors representation). This figure illustrates the velocity-field of  two CFH2021 \cite{dan2021price} treatments. Using price-strategy game data from six participants \cite{dan2021price}, we discretize the continuous strategy space by tertiles (i.e., Rock-Paper-Scissors ). Specifically, the discretization in this figure adopts a fixed-range uniform tertile partition. The lower and upper bounds are taken from Table~1 of  CFH2021 \cite{dan2021price}, namely the minimum and maximum NE prices for each treatment: for the Unstable Discrete Time treatment (panel (a)), the range is $[0.9, 1.8]$; for the Stable Discrete Time treatment (panel (b)), the range is $[1.11, 1.8]$. Each range is divided into three equal intervals, yielding the three price levels (Low, Middle, High). Consequently, the six players' strategies, originally continuous observations, are converted into choices among three strategies $(R,P,S)$, which yields an equivalent structure of a six-player Rock-Paper-Scissors game \cite{wang2013,wang2014social,wang2011rps}. All observable social states are thus compressed into 28 combinatorial states. These 28 states can be represented on a simplex lattice $G^{28}$ in the three-dimensional strategy space, where each state corresponds to a three-dimensional vector $x_i = (p_R, p_P, p_S)$, with $n_R$ denoting the proportion of players choosing $R$ among the six (though the notation here uses $p_R$ for the count, the proportion is $n_R/6$; we retain the original notation). This representation is consistent with the framework featured as MIT Technology Review's Best of 2014 \cite{wang2014social, MIT2014Best,wang2011rps}. Using the velocity-field approach (central-difference estimation of local dynamical trends at each state, see \cite{wang2017}), we obtain an empirical velocity vector field defined on the discrete simplex lattice $G^{28}$, which characterizes the direction and magnitude of strategy evolution. Referring to velocity definition in Eq. (\ref{eq:velocity_exp}), in a one-dimensional projection of  the manifold, the geometric distance weighting factor is a scalar $(j-i)$, determined by the difference between two social states along that single dimension. In a two-dimensional projection, the geometric distance weighting factor is a two-dimensional vector $\mathbf{x}_j - \mathbf{x}_i$, because $\mathbf{x}$ is a two-dimensional vector. The figure above shows the experimental velocity fields obtained from the CFH2021 \cite{dan2021price} data for the Stable-Discrete Time treatment and
Unstable-Discrete Time Treatment, computed using Eq. (\ref{eq:velocity_exp}) with the experimental manifolds. The \textbf{size of bubbles} at the lattices reflects the frequency being visited in the treatment in total. The original experimental data consist of 1200 observations of rounds for the Unstable treatment and 1198 for the Stable treatment . Referring to the velocity definition Eq.(\ref{eq:velocity_exp}) based on consecutive rounds, the aggregate frequency of states used in the velocity calculation amounts to 1998 (Stable) and 1996 (Unstable).
%原始实验数据中，稳定处理包含1,198条观测记录，不稳定处理包含1,200条。依据基于相邻轮次的速度定义（式\ref{eq:velocity_exp}），实际用于速度场计算的状态频次总和分别为稳定处理1,998、不稳定处理1,996。}
    \label{fig:RPS-dan2021-D} }
\end{figure}

\begin{figure}[!ht]
    \centering
    
    \begin{subfigure}[b]{0.80\linewidth}
        \centering
 \caption{Theoretical eigencycle spectrum ($\vartheta^T$)}
        \label{fig:eicycle_dyn_main}       \includegraphics[width=\linewidth]
        {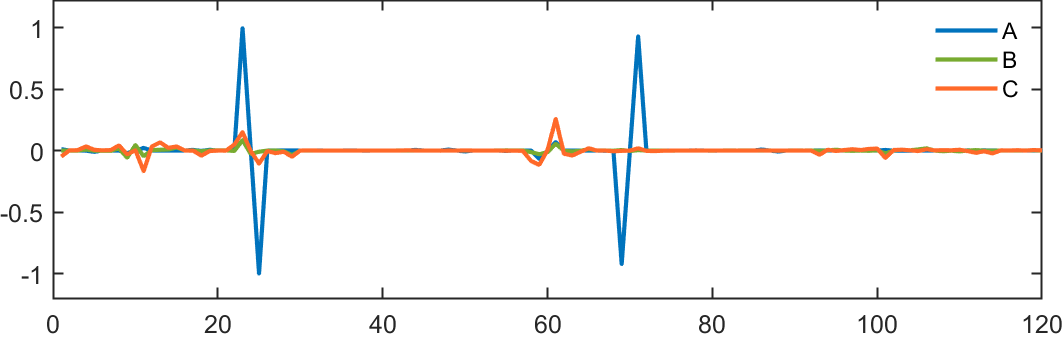}
        
    \end{subfigure}
    % \hfill
    \begin{subfigure}[b]{0.80\linewidth}
        \centering
 \caption{Experimental eigencycle spectrum($\vartheta^E$)}
        \label{fig:exp_3eigencycle}       \includegraphics[width=\linewidth]
        {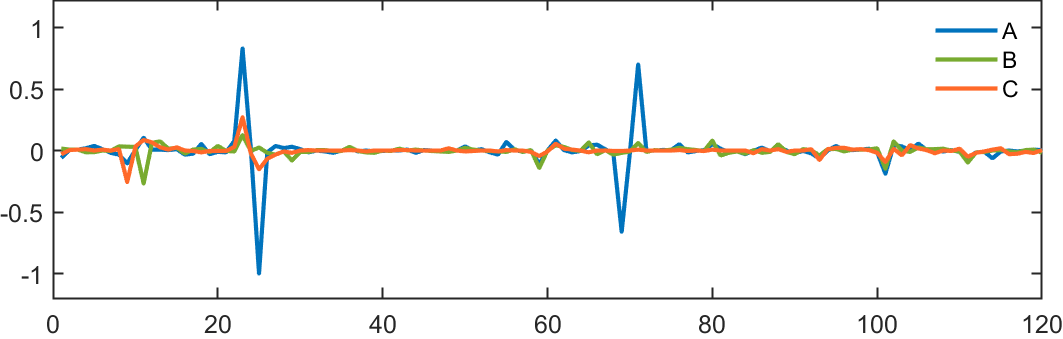}
        
    \end{subfigure}
\caption{Eigencycle spectra of the game. The parameters for treatments A, B, and C are $(m,n)=(2,1)$, $(3,2)$, and $(4,2)$, respectively. (a)~Theoretical spectrum $\vartheta^{T}$ predicted by the game dynamics. (b)~Experimental spectrum $\vartheta^{E}$ measured from human-subject experiments. Both spectra are plotted on the same ordering of two-dimensional subspaces $\Omega^{mn}$ defined by Eq.~\eqref{eq:mn_order} on the horizontal axis (This is a two-player asymmetric game. Each player has 8 strategies, so the fully expanded strategy space is 16-dimensional. Therefore, according to the definition of the eigenmanifold, there are 120 measurable two-dimensional subspaces 
$(m,n)$. Hence, the horizontal axis is labeled by the integers from 1 to 120). The normalized eigenmanifold value at each ($m,n$) subspace on the vertical axis. So that theory and experiment can be compared componentwise. The experiment aims to investigate the collapse process in the full Nash equilibrium finding process and to test the completeness and consistency of the game dynamics paradigm \cite{WW2023pulse}.}
    \label{fig:eigencycle_spectra}
\end{figure}

% \begin{verbatim}
%  本征流形的二维速度场表示(石头剪刀布表示). 这里是速度场算式的举例. 这是通过6个参与者选择的价格策略博弈数据 \cite{dan2021price}, 对连续策略空间进行3分位数离散化(也就是石头剪刀布化), \textcolor{blue}{具体而言，本图中的离散化采用固定边界的均匀三等分,上下边界取自 CFH2021 \cite{dan2021price} 中 Table~1 的 NE 价格最小值和最大值：不稳定离散时间处理组（子图 (a)）的范围为 $[0.9, 1.8]$，稳定离散时间处理组（子图 (b)）的范围为 $[1.1, 1.8]$。每个范围等分为三个区间，从而得到三个价格水平（低、中、高）。}这样6人的策略从原来的连续观测值转化为在三策略$(R,P,S)$中进行选择, 这就成为6人参与的石头剪刀布博弈的等价结构, 也就是所有可观测的社会态压缩至 28 个组合状态. 
% 这28个态可以在三维策略空间上表示为一个单纯形点阵$G^{28}$, 每个态都对应着一个三维的空间向量 $x_i = (p_R,p_p,p_s)$, 其中, $n_R$表示6人中使用 $R$策略的比例.  这和被MIT Technology Review推荐为2014年度最佳的(Best of 2014的石头剪刀布的表示方法一致\cite{wang2014social,MIT2014Best,wang2011rps}) 的框架图.利用速度场表示(中心差分估计每个状态上的局部动力学趋势, 见\cite{wang2017})，最终获得一个定义在离散单纯形点阵$G^{28}$上的经验速度向量场，用于刻画策略演化的方向和大小的值。在流形在一维速度场的投影中, 几何空间距离权重因子是一维向量$(j-i)$, 取决于二个社会态之间的一维的差值. 在二维速度场的投影中, 几何空间距离权重因子是二维向量, 是$x_j - x_i$, 这是因为$x$是一个二维向量. 上图是 CFH2021, Stable-Discrete Time Treatment 和 Unstable-Discrete Time Treatment 的数据, 采用算式 \ref{eq:velocity_exp} 获得的实验速度场的结果. ~~~~   
% \end{verbatim}

% \begin{figure}
%     \centering
%     \includegraphics[width=0.45\linewidth]{Dan2021-SD-RPS.png}
%     \includegraphics[width=0.45\linewidth]{Dan2021-UD-RPS.png}

% \end{figure}

% \section{附录方法\label{sec:append}}
\section{Appendix Method\label{sec:append}}

% \subsection{理论本征流形的定义与算法\label{sec:eigenmanifold}}
\subsection{Definition of Theoretical Manifolds\label{sec:eigenmanifold}}

% \subsubsection{前提条件\label{Prerequisites}}
\subsubsection{Prerequisites\label{Prerequisites}}

% 对于给定的博弈，博弈动力学系统可以表示为一个速度向量场
% \cite{2011Sandholm}\cite{dan2016}：

 For a given game, the game dynamics system can be represented as a velocity vector field
\cite{2011Sandholm}\cite{dan2016}:

\begin{equation}
    \dot{x} = f(x) 
\end{equation}

% 其中 $x\in R^N$，$N$ 为策略空间的维数。
%
%  纳什均衡是向量场的一个均衡点(满足 $\dot{x} = 0$ 的点)。
% 在均衡点附近的线性近似下，动力学可以表示为 
% \cite{2011Sandholm}\cite{dan2016}：
%

 where $x\in R^N$, $N$ is the dimension of the strategy space.
A Nash equilibrium is an equilibrium point of the vector field (a point satisfying $\dot{x} = 0$).
Under a linear approximation near the equilibrium point, the dynamics can be expressed as 
\cite{2011Sandholm}\cite{dan2016}:

\begin{equation}
  \dot{x} = J  x  
\end{equation}

%
% 其中 $J$ 是均衡点处的雅可比矩阵(本征矩阵)。

 %
where $J$ is the Jacobian matrix (characteristic matrix) at the equilibrium point.

% 假设 $\xi_i$ 是与可对角化矩阵 $J$ 的本征值 $\lambda_i$
% 相关联的本征向量；并假设初始条件可以表示为
%    $x(0) = \sum_{i=1}^N a_i {\xi}_i$，
% 那么演化轨迹可以表示为
%

 Assume $\xi_i$ is the eigenvector associated with the eigenvalue $\lambda_i$ of the diagonalizable matrix $J$; and assume the initial condition can be expressed as
   $x(0) = \sum_{i=1}^N a_i {\xi}_i$,
then the evolution trajectory can be expressed as

\begin{equation}\label{eq:eigdeco2}
x(t) = \sum_{i=1}^N e^{\lambda_it} a_i  {\xi}_i.   
\end{equation}

%
% 此处，本征向量 $\xi_i$ 描述了一个本征模态(振荡系统中的简正模态，
% 可能包含多个分量)，其中所有分量都以相同的频率 $\lambda_i$ 振荡。
% %
% 只有\textbf{具有非零虚部的本征值}(即复本征值)对应的本征
% 向量才会产生周期运动；实本征值对应指数增长/衰减，不产生循环。

 %
Here, the eigenvector $\xi_i$ describes a characteristic mode (a normal mode in an oscillatory system, which may contain multiple components), where all components oscillate at the same frequency $\lambda_i$.
Only eigenvectors corresponding to \textbf{eigenvalues with non-zero imaginary parts} (i.e., complex eigenvalues) produce periodic motion; real eigenvalues correspond to exponential growth/decay and do not generate cycles.

% \subsubsection{定义本征流形向量和它的分量本征圈标量\label{sec:Defining_Eigenmanifold}}
\subsubsection{Defining the Eigenmanifold Vector and Its Component Eigen-Circle Scalars\label{sec:Defining_Eigenmanifold}}

% 参照\cite{WY2020,Wang2022A4}每个本征流形$\sigma_k$由本征向量 $\xi_k$中的元素决定. 对于一个
% 本征向量 $\xi_k$ 中的任意两个分量 $\eta_m$ 和 $\eta_n$
% ($m \neq n$)，

 Refer to \cite{WY2020,Wang2022A4}, each eigenmanifold $\sigma_k$ is determined by the elements in the eigenvector $\xi_k$. For any two components $\eta_m$ and $\eta_n$ ($m \neq n$) in an eigenvector $\xi_k$,

\begin{equation}
    \xi_k = (\eta_1...\eta_m...\eta_n...,\eta_N)^T
\end{equation}

% 定义它们在二维子空间$\Omega^{mn}, m,n \in \{1,...,N\}$上的\textbf{本征圈值} 为：

Define their \textbf{eigencircle values} on the two-dimensional subspace $\Omega^{mn}, m,n \in \{1,...,N\}$ as:

\begin{equation}\label{eq:theo_sigma_mn}
\sigma^{mn} = \pi \cdot \|\eta_m\| \cdot \|\eta_n\| 
\cdot \sin\bigl(\arg(\eta_m) - \arg(\eta_n)\bigr) 
\end{equation}

% 其中： 
% $\|\eta_m\|$ 是复数 $\eta_m$ 的模(振幅)，
% $\arg(\eta_m)$ 是 $\eta_m$ 的辐角(相位)。

 where: 
$\|\eta_m\|$ is the modulus (amplitude) of the complex number $\eta_m$,
$\arg(\eta_m)$ is the argument (phase) of $\eta_m$.

%

% \paragraph{本征流形向量$\sigma_k$ }
\paragraph{Eigenmanifold $\sigma_k$ vector and the order of its components}

% 本征流形向量$\sigma_k$ 由本征圈$\sigma^{mn}$组成, 顺序为

 The eigenmanifold vector $\sigma_k$ is composed of eigencycles $\sigma^{mn}_k$, in the order, namely \textbf{ Order of Manifold Vector Components}

\begin{equation} \label{eq:mn_order}
 \sigma_k = 
\Big(\sigma_k^{1,2},\sigma_k^{1,3},...,\sigma_k^{1,N},,\
\sigma_k^{2,3},..,\sigma_k^{2,N},...,\sigma_k^{N-
1,N}\Big)^T 
 \end{equation}

% 也就是$\sigma_k^{i,j}$中. 先$i$从$1,2...N-1$排序, 对应每一个
% 给定的$i$, $j$从$i+1$到$N$排序, 其中的每个元素
% $\sigma_k^{m,n}$由$\xi_k$中的($m,n$)分量决定, 见.公式
% (\ref{eq:mn_order}). 显然有

 That is, in $\sigma_k^{i,j}$. First, $i$ is ordered from $1, 2, \dots, N-1$, and for each given $i$, $j$ is ordered from $i+1$ to $N$, where each element $\sigma_k^{m,n}$ is determined by the $(m,n)$ component in $\xi_k$, see formula (\ref{eq:mn_order}). Obviously,

% \begin{itemize}
%     \item 该向量的元素个数为 $N(N-1)/2$
%     \item 对于实本征值而言, 本征流形向量为 $0$ 向量. 
%     \item 对于二个互为共轭的复本征向量, 其对应的本征流形向量是互
%     为反数. 
% \end{itemize}

\begin{itemize}
    \item The number of elements of this vector is $N(N-1)/2$.
    \item For real eigenvalues, the eigenmanifold vector is the zero vector.
    \item For two complex eigenvectors that are conjugates of each other ($\xi_k$ and $\xi_{k^+}$), their corresponding eigenmanifold vectors are opposites of each other, $\sigma^{mn}_k$=$-\sigma^{mn}_{k^+}$.
    \item The sign convention is fixed by taking the member of the conjugate pair with $\text{Im} \lambda_k > 0$ 
\end{itemize}

\paragraph{Eigenmanifold Vector normalization}
Since the overall amplitude of an eigenmanifold vector depends on the arbitrary scaling of the eigenvector $\xi_k$, the eigenmanifold vectors are reported in the normalized form
\begin{equation}\label{eq:sigma_norm}
\tilde{\sigma}_k = \frac{\sigma_k}{\lVert\sigma_k\rVert_2},
\qquad
\lVert\sigma_k\rVert_2 = \Big(\sum_{m<n}\bigl(\sigma_k^{mn}\bigr)^2\Big)^{1/2},
\end{equation}
where the same normalization factor is applied to the theoretical and the experimental vectors when the two are compared componentwise.

% \paragraph{本征流形向量空间$\mathcal{V}$}

\paragraph{Eigenmanifold Vector Space~$\mathcal{V}$}

% 由 $\sigma_k(k\in (1,...,N))$构成的基向量而展开的抽象空间,叫本
% 征流形向量空间 $\mathcal{V}$,
% $$\mathcal{V} = \bigoplus_{k=1}^{N} \operatorname{span}\{\sigma_k\}$$
% 本文所验证的"博弈稳态系统的动力学结构是由本征流形线性叠加而成"的理
% 论假设, 就是指, 稳态博弈运动可以在这个抽象空间的$\sigma_k$上线性
% 展开. 展开中, 权重最大的项是主导项, 本文取特征值复数部分最大的
% 本征值所对应的本征流形作为主导项, 作为所有失真系统的直接理论预估.

 The abstract space expanded by the basis vectors formed by theoretical eigenmanifold vector $\sigma_k(k\in (1,...,N))$ is called the eigenmanifold vector space $\mathcal{V}$,
$$\mathcal{V} = \bigoplus_{k=1}^{N} \operatorname{span}\{\sigma_k\}$$
The theoretical hypothesis verified in this paper—that the dynamical structure of a game steady-state system is composed of a linear superposition of eigenmanifolds, means that, steady-state game motion can be linearly expanded on $\sigma_k$ in this abstract space. 

In the expansion, the term with the largest weight is the principal term, which is related to the eigenvalue with max imaginary and the weight ordering is of various mode in an ideal model being derived in Appendix~\ref{sec:app_dom}.

\paragraph{From the trajectory decomposition to the manifold expansion\label{pra:equilvant2obj}}
The hypothesis of Eq.~(\ref{eq:main}) is stated for the manifold vectors, whereas the eigen decomposition of the linearized dynamics is stated for the state vector, Eq.~(\ref{eq:eigdeco2}); at first sight the two objects are not the same Following \cite{WY2020,2021Qinmei,Wang2022A4},  we explain why they are equivalent when cyclic motion is tested.\\

The experimental manifold vector is a time average of a bilinear functional of the trajectory, Eq.~(\ref{eq:experiment_sigma_mn_L}), so inserting the modal expansion of $x(t)$ into $L^{mn}(t)$ produces, besides the diagonal (single-mode) terms, cross terms between different modes exist. \\

Writing the deviation from the equilibrium as $\dot{x}(t)=\sum_k c_k(t)\xi_k$, the diagonal term of mode $k$ contributes $w_k\sigma^{mn}_k$ with $w_k \propto \omega_k$ and $\omega_k\propto \operatorname{Im}\lambda_k\rvert$, whereas each cross term oscillates at the difference frequency is therefore suppressed. As an approximate, we ignore the bias from orthogonal of the $\xi$-set as base at current  stage. \\

Since the frequencies $\omega_k$ are distinct and the average runs over the whole record, the cross terms are negligible and the expansion of Eq.~(\ref{eq:main}) is recovered. This is the step that licenses testing the hypothesis on the manifold vectors; the argument is completed in the simplified model shown in Appendix~\ref{sec:app_dom}, paragraphs (4)--(5). 

% \paragraph{ 本征向量及其本征值、本征流形的算例}
\paragraph{Example of Eigenvectors, Eigenvalues, and Eigenmanifolds}

\begin{longtable}{|r|c|c|c|c|c|}
\caption{Theoretical Values of Eigenvectors, Eigenvalues, and Eigenmanifolds}
\label{tab:theory_value_eign}\\
\hline
\textbf{eigenvector} & $\eta_1$ &  $\eta_2$  &  $\eta_3$  &  $\eta_4$  &  $\eta_5$  \\
\hline
\endhead
\hline
\multicolumn{6}{|c|}{Continuous Table} \\
\hline
\endfoot
\hline
\endlastfoot
-component 1 & $-0.798$ & $-0.798$ & $-0.911$ & $0.834$ &
$-0.843$ \\
$2$ & $0.288-0.331i$ & $0.288+0.331i$ & $-0.356$ & 
$-0.416$ & $0.284$ \\
$3$ & $0.325+0.083i$ & $0.325-0.083i$ & $-0.184$ &
$0.060$ & $0.069$ \\
$4$ & $0.145+0.177i$ & $0.145-0.177i$ & $-0.093$ & 
$-0.319$ & $0.040$ \\
$5$ & $0.039+0.070i$ & $0.039-0.070i$ & $-0.029$ & 
$-0.159$ & $0.450$ \\
\hline
\textbf{eigenvalue} &$\lambda_1$&$\lambda_2$&$\lambda_3$& $\lambda_1$& $\lambda_5$\\
& $-2.936+3.204i$ & $-2.936-3.204i$ & $0$ &
$-2.341$ & $-0.918$ \\
\hline
\textbf{eigenmanifold} &$\sigma_1$&$\sigma_2$&$\sigma_3$&$\sigma_4$&$\sigma_5$\\
$1~~~~ (1,2)$ & $-0.828$ & $0.828$ & $0$ & $0$ & $0$ \\
$2~~~~ (1,3)$ & $0.209$ & $-0.209$ & $0$ & $0$ & $0$ \\
$3~~~~ (1,4)$ & $0.444$ & $-0.444$ & $0$ & $0$ & $0$ \\
$4~~~~ (1,5)$ & $0.176$ & $-0.176$ & $0$ & $0$ & $0$ \\
$5~~~~ (2,3)$ & $-0.413$ & $0.413$ & $0$ & $0$ & $0$ \\
$6~~~~ (2,4)$ & $-0.311$ & $0.311$ & $0$ & $0$ & $0$ \\
$7~~~~ (2,5)$ & $-0.104$ & $0.104$ & $0$ & $0$ & $0$ \\
$8~~~~ (3,4)$ & $-0.143$ & $0.143$ & $0$ & $0$ & $0$ \\
$9~~~~ (3,5) $& $-0.061$ & $0.061$ & $0$ & $0$ & $0$ \\
$10~~~ (4,5)$ & $-0.010$ & $0.010$ & $0$ & $0$ & $0$ \\
\end{longtable}

\normalsize

% \textcolor{red}{添加一个算例, 根据本文中公式, 详细给出}\\
% 以$\sigma_k^{1,5}=0.176$的计算为例：

% \textcolor{red}{Add an example, and provide details based on the formulas in this paper.}\\
Take the calculation of $\sigma_k^{1,5}=0.176$ as an example:

% \begin{itemize}
%     \item $\sigma_k^{1,5}$对应本征圈$\sigma^{15}$。
%     \begin{itemize}
%         \item  $m=1$对应的本征向量分量：$\eta_1=−0.798$
%         \item $n=5$对应的本征向量分量：
%         $\eta_5=0.039+0.070i$
%     \end{itemize}
   
%     \item 计算模(振幅)：
%     \begin{itemize}
%         \item $\|\eta_1\|=|-0.798|=0.798$
%         \item  $\|\eta_5\|=\sqrt{0.039^2+0.070^2}=
%         \sqrt{0.006421} \approx 0.08013$
%     \end{itemize}
    
%     \item 计算辐角(相位)：
%     \begin{itemize}
%         \item $\arg(\eta_1)=\pi$
%         \item  $\arg(\eta_5)=\arctan\left(\frac{0.070}
%         {0.039}\right)
%         =\arctan\left(\frac{70}{39}\right)$
%         \item $\sin\bigl(\arg(\eta_1) - 
%         \arg(\eta_5)\bigr)=\sin\bigl(\pi - 
%         \arctan\frac{0.070}
%         {0.039}\bigr)=\sin\bigl( 
%         \arctan\frac{0.070}
%         {0.039}\bigr)=\frac{0.070}{\|\eta_5\|} \approx 0.8736$
%     \end{itemize}   

%     \item 代入公式(\ref{eq:theo_sigma_mn})求解：
    
% \begin{equation}
% \begin{aligned}
% \sigma^{1,5} &= \pi \cdot \|\eta_1\| \cdot \|\eta_5\|
% \cdot \sin\bigl(\arg(\eta_1) - \arg(\eta_5)\bigr) \\
% &= 3.1416 \times 0.798 \times 0.08013 \times 0.8736 \\
% &\approx 0.176
% \end{aligned}
% \end{equation}

% \end{itemize}

\begin{itemize}
    \item $\sigma_k^{1,5}$ corresponds to the eigencircle $\sigma^{1,5}$.
    \begin{itemize}
        \item  Eigenvector component for $m=1$: $\eta_1=-0.798$
        \item Eigenvector component for $n=5$:
        $\eta_5=0.039+0.070i$
    \end{itemize}
   
    \item Compute the modulus (amplitude):
    \begin{itemize}
        \item $\|\eta_1\|=|-0.798|=0.798$
        \item  $\|\eta_5\|=\sqrt{0.039^2+0.070^2}=
        \sqrt{0.006421} \approx 0.08013$
    \end{itemize}
    
    \item Compute the argument (phase):
    \begin{itemize}
        \item $\arg(\eta_1)=\pi$
        \item  $\arg(\eta_5)=\arctan\left(\frac{0.070}
        {0.039}\right)
        =\arctan\left(\frac{70}{39}\right)$
        \item $\sin\bigl(\arg(\eta_1) - 
        \arg(\eta_5)\bigr)=\sin\bigl(\pi - 
        \arctan\frac{0.070}
        {0.039}\bigr)=\sin\bigl( 
        \arctan\frac{0.070}
        {0.039}\bigr)=\frac{0.070}{\|\eta_5\|} \approx 0.8736$
    \end{itemize}   

    \item Substitute into formula (\ref{eq:theo_sigma_mn}) to solve:
    
\begin{equation}
\begin{aligned}
\sigma^{1,5} &= \pi \cdot \|\eta_1\| \cdot \|\eta_5\|
\cdot \sin\bigl(\arg(\eta_1) - \arg(\eta_5)\bigr) \\
&= 3.1416 \times 0.798 \times 0.08013 \times 0.8736 \\
&\approx 0.176
\end{aligned}
\end{equation}

\end{itemize}

% \subsection{实验流形值的测量}\label{sec:exp_L}
\subsection{Measurement of Experimental Manifold Vector }\label{sec:exp_L}

% 根据《ONeill\_human\_game》论文(Wang \& Yao, 2021)的第3.2
% 节、附录6.2以及附录6.5， 从实验时间序列中计算实验流向量 $\bar{L}$的方
% 法总结如下：

 According to Section 3.2, Appendix 6.2, and Appendix 6.5 of the paper "ONeill\_human\_game" (Wang \& Yao, 2021 \cite{2021Qinmei,WY2020}), the method for calculating the experimental manifold vector $L$ from the experimental time series is summarized as follows:

% \subsubsection{基本定义\label{sec:BasicDefin}}
\subsubsection{Basic Definition\label{sec:BasicDefin}}
% 实验流向量是一个向量$L$, 由$N(N-1)/2$构成. 每个元素是对应的二维
% 子空间上测量到的实验子空间角动量$L^{mn}$. 实验子空间角动量用于测量在
% 二维子空间 $\Omega^{mn}$ 中，两个状态变量 $p_m(t)$ 和 
% $p_n(t)$ 随时间的\textbf{运动强度}。它是瞬时有向面积变化率
% 的一种度量。

The experimental manifold vector is a vector $L$, composed of $N(N-1)/2$ elements. Each element corresponds to the angular momentum, $L^{mn}$ measured in a two-dimensional subspace. Angular momentum is used to measure the \textbf{cyclic motion intensity} of two state variables $p_m(t)$ and $p_n(t)$ over time in the two-dimensional subspace $\Omega^{mn}$. It is a measure of the instantaneous rate of change of directed area.

% 实验数据通常是离散时间步(例如每轮或每秒记录一次状态)。对于离散时间
% 序列 $t = 1, 2, \dots, T$，可以用\textbf{差分近似}导数：

 Experimental data are typically recorded at discrete time steps (e.g., recording the state every round or every second). For a discrete time
series $t = 1, 2, \dots, T$, the derivative can be approximated using \textbf{finite differences}:

\begin{equation}
 \frac{d p_m}{dt}(t) = \frac{p_m(t+\Delta t) - p_m(t)}
 {\Delta t}   
\end{equation}

% 这里， 社会状态向量分量 $p_i(t)$   表示$t$选择 $i$ 纯策略的玩家
% 比例。若时间间隔 $\Delta t $均匀且归一化为单位间隔(如1轮)，
% 则简化为：

 Here, the social state vector component $p_i(t)$ represents the proportion of players choosing pure strategy $i$ at time $t$. If the time interval $\Delta t$ is uniform and normalized to a unit interval (e.g., 1 round), it simplifies to:

\begin{equation}\label{eq:experiment_sigma_mn_L}
L^{mn}(t) = p_m(t) \cdot \bigl(p_n(t+1) - 
p_n(t)\bigr) - p_n(t) \cdot \bigl(p_m(t+1) - 
p_m(t)\bigr) 
\end{equation}

% 为了获得稳健的统计量，通常全部实验的平均实验流形向量：

 To obtain robust statistics, the average angular momentum of all experiments is typically used:

\begin{equation}\label{eq:bar_L_mn}
 \bar{L^{mn}} = \frac{1}{T-1} \sum_{t=1}^{T-1} L^{mn}
 (t)   
\end{equation}

% 其中$T$是总时间轮次. 在统计检验公式(\ref{eq:main})中, 
% $\bar{L}$是$L^{mn}$ 按照 表达式 (\ref{eq:mn_order}) 的顺序
% 构成的向量, 也就是

 where $T$ is the total number of time rounds. In the statistical test formula (\ref{eq:main}), 
$\bar{L}$ is the vector formed by $L^{mn}$ according to the order of expression (\ref{eq:mn_order}), 
that is,

\begin{equation} \label{eq:L_order}
 L = 
 \Big(L^{1,2},L^{1,3},...,L^{1,N},,L^{2,3},..,L^{2,N},..
 .,L^{N-1,N}\Big)^T 
 \end{equation}

% 显然有

Obviously, there is

% \begin{itemize}
%     \item 该向量的元素个数为 $N(N-1)/2$, 和理论计算中的本征流形
%     向量(定义见公式\ref{eq:mn_order})一样, 
%     \item $L$向量的子空间指标顺序和见公式\ref{eq:mn_order})中
%     本征流形 $\sigma_k$的顺序也一样.   
% \end{itemize}

\begin{itemize}
    \item The number of elements in this vector is $N(N-1)/2$, which is the same as the theoretical eigenmanifold vector in (see definition in equation \ref{eq:mn_order}).
    \item The order of the subspace indices of the $L$ vector is the same as the order of the eigenmanifold $\sigma_k$ in equation \ref{eq:mn_order}.
\end{itemize}

% 这是实验值可能可以在本征流形上展开的基础.

 This is the basis on which experimental values may be expanded on the eigenmanifold.

% \subsubsection{本征圈实验测量的例子\label{sec:exp_measure_examlp}}
\subsubsection{Experimental Subspace Angular Momentum  $\bar{L}^{mn}$ \label{sec:exp_measure_examlp}}

% \textcolor{red}{添加一个算例, 根据本文中测量公式, 逐个符号对
% 应, 详细给出}

\begin{table}[!ht]
    \centering
    \begin{tabular}{c|c|c|c|c|c}
    \hline
Time($t$) &$p_1$    &$p_2$  &$p_3$  &$p_4$  &$p_5$  \\ \hline
1&	0.545&	0.182&	0.182&	0&	0.091\\ 
2&	0.455&	0.182&	0.182&	0&	0.182\\ 
3&	0.273&	0&	0.455&	0&	0.273\\ 
4&	0.182&	0&	0.455&	0&	0.364\\ 
5&	0&	0.091&	0.727&	0&	0.182\\ 
6&	0&	0.182&	0.727&	0&	0.091\\ \hline
    \end{tabular}
    \caption{Experiment Data from Dan2026all,$\gamma=2$}%[4.198,10]分割成5段
    \label{tab:placeholder}
\end{table}
% 以$L^{1,5}=0.043$的计算为例：

Take the calculation of $L^{1,5} = 0.043$ as an example:

% \begin{itemize}
%     \item 示例计算：$L^{(1,5)}(t=2)$.取状态变量$p_1(t),~p_5(t);~p_1(t+1),~p_5(t+1)$.
%     \begin{itemize}
%         \item  $t=2$时，$p_1(2)=0.455$ ~~$p_5(2)=0.182$
%         \item $t=3$时，$p_1(3)=0.273$ ~~ $p_5(3)=0.273$
%     \end{itemize}
   
%     \item 代入公式(\ref{eq:experiment_sigma_mn_L})计算实验子空间角动量值：
%     \begin{equation}
%     \begin{aligned}
% L^{(1,5)}(2) &=p_1(2) \cdot \bigl(p_5(3) - 
% p_5(2)\bigr) - p_5(2) \cdot \bigl(p_1(3) - 
% p_1(2)\bigr)  \\
% &= 0.455 \cdot \bigl(0.273 - 
% 0.182\bigr) - 0.182 \cdot \bigl(0.273 - 
% 0.455\bigr)\\
% &\approx 0.075
% \end{aligned}
% \end{equation}
    
%     \item 相同方法计算得到：
%     \begin{itemize}
%         \item $L^{(1,5)}(1)=0.058$
%         \item $L^{(1,5)}(3)=0.050$
%         \item $L^{(1,5)}(4)=0.033$
%         \item $L^{(1,5)}(5)=0$
%     \end{itemize}

%     \item 代入公式(\ref{eq:bar_L_mn})求解：
    
% \begin{equation}
% \begin{aligned}
% L^{1,5}&= \bar{L^{1,5}} = \frac{1}{5} \sum_{t=1}^{5} L^{(1,5)}(t) = \frac{1}{5} \Big(L^{(1,5)}(1)+L^{(1,5)}(2)+L^{(1,5)}(3)+L^{(1,5)}(4) +L^{(1,5)}(5) \Big) \\
% &= \frac{1}{5}\times \Big(0.058+0.075+0.050+0.033+0 \Big)\\
% &=0.043
% \end{aligned}
% \end{equation}

% \end{itemize}

\begin{itemize}
    \item Example calculation: $L^{(1,5)}(t=2)$. Take state variables $p_1(t),~p_5(t);~p_1(t+1),~p_5(t+1)$.
    \begin{itemize}
        \item At $t=2$, $p_1(2)=0.455$ ~~ $p_5(2)=0.182$
        \item At $t=3$, $p_1(3)=0.273$ ~~ $p_5(3)=0.273$
    \end{itemize}
   
    \item Substitute into equation (\ref{eq:experiment_sigma_mn_L}) to calculate the angular momentum value:
    \begin{equation}
    \begin{aligned}
L^{(1,5)}(2) &=p_1(2) \cdot \bigl(p_5(3) - 
p_5(2)\bigr) - p_5(2) \cdot \bigl(p_1(3) - 
p_1(2)\bigr)  \\
&= 0.455 \cdot \bigl(0.273 - 
0.182\bigr) - 0.182 \cdot \bigl(0.273 - 
0.455\bigr)\\
&\approx 0.075
\end{aligned}
\end{equation}
 Using the same method, we obtain: 
          $L^{(1,5)}(1)=0.058$, 
       $L^{(1,5)}(3)=0.050$,
       $L^{(1,5)}(4)=0.033$,
       $L^{(1,5)}(5)=0$. 

    \item Substitute into equation (\ref{eq:bar_L_mn}) to solve:
    
\begin{equation}
\begin{aligned}
L^{1,5}&= \bar{L^{1,5}} = \frac{1}{5} \sum_{t=1}^{5} L^{1,5}(t) = \frac{1}{5} \Big(L^{(1,5)}(1)+L^{(1,5)}(2)+L^{(1,5)}(3)+L^{(1,5)}(4) +L^{(1,5)}(5) \Big) \\
&= \frac{1}{5}\times \Big(0.058+0.075+0.050+0.033+0 \Big)\\
&=0.043
\end{aligned}
\end{equation}

\end{itemize}

\subsection{Table of the Statistical Test Results\label{sec:Statis_table}}
\subsubsection{F-test for Result 1,2}
% \subsubsection{F-test}
See Table \ref{tab:F-test_full}, Table \ref{tab:F-test_1st}
% , Table \ref{tab:F-test_M1st}, Table \ref{tab:F-test_M2nd} 
and description in the main text in section \ref{sec:statis_Test_result}.

\begin{table}[!ht]
\centering
\caption{Statistical F-Test Results: Full model regression (Full)}
\label{tab:F-test_full}
\resizebox{\textwidth}{!}{%
\begin{tabular}{c | c| c| c| c| c|c}
\hline
Treatment  & CFH2021 & CFH2021 & CFGS2026 &
CFGS2026 & CFGS2026 & CFGS2026
\\ 
 & UD & SD & $\gamma2$ &
$\gamma3.6$ & $\gamma4.4$ & $\gamma6$
\\\hline
 bin\_num & $F_1$ & $F_2$ & $F_3$  & 
 $F_4$ &$F_5$  & $F_6$ \\
 & ($p_1$) & ($p_2$)& ($p_3$) & 
 ($p_4$)&($ p_5$) & ($p_6$)\\
\hline
% ========== Full 部分 ==========
\multirow{2}{*}{ 10}&	\makecell[t]{11 \\(4.22E-06)} &	\makecell[t]{17 \\(3.17E-08)} &	\makecell[t]{6.49 \\(1.07E-03)} &	\makecell[t]{4.55 \\(7.69E-03)} &	\makecell[t]{20.1 \\(3.60E-08)} &	\makecell[t]{59.3 \\(5.74E-15)} \\
\multirow{2}{*}{20}&	\makecell[t]{13.8 \\(8.23E-17)} &	\makecell[t]{26.3 \\(1.48E-28)} &	\makecell[t]{8.91 \\(2.01E-09)} &	\makecell[t]{5.66 \\(6.33E-06)} &	\makecell[t]{16.2 \\(1.75E-16)} &	\makecell[t]{63.8 \\(1.06E-45)} \\
\multirow{2}{*}{30}&	\makecell[t]{15.3 \\(5.29E-30)} &	\makecell[t]{36.2 \\(2.70E-63)} &	\makecell[t]{13.2 \\(1.08E-21)} &	\makecell[t]{7.4 \\(1.23E-11)} &	\makecell[t]{12.4 \\(1.99E-20)} &	\makecell[t]{73.3 \\(8.69E-91)} \\
\multirow{2}{*}{40}&	\makecell[t]{15.5 \\(1.60E-42)} &	\makecell[t]{37.6 \\(6.75E-96)} &	\makecell[t]{19.5 \\(2.95E-42)} &	\makecell[t]{9.45 \\(1.31E-20)} &	\makecell[t]{10.9 \\(2.57E-24)} &	\makecell[t]{78.4 \\(3.14E-143)} \\
\multirow{2}{*}{50}&	\makecell[t]{14.5 \\(2.73E-51)} &	\makecell[t]{42.9 \\(9.53E-143)} &	\makecell[t]{23.3 \\(7.85E-69)} &	\makecell[t]{11.6 \\(5.35E-33)} &	\makecell[t]{9.41 \\(7.70E-26)} &	\makecell[t]{86.6 \\(1.71E-209)} \\
\multirow{2}{*}{60}&	\makecell[t]{16.7 \\(5.92E-73)} &	\makecell[t]{40.3 \\(7.68E-171)} &	\makecell[t]{28.7 \\(1.56E-100)} &	\makecell[t]{12.5 \\(2.34E-45)} &	\makecell[t]{8.38 \\(4.67E-27)} &	\makecell[t]{93.4 \\(8.81E-284)} \\
\multirow{2}{*}{70}&	\makecell[t]{18.6 \\(1.27E-97)} &	\makecell[t]{42.9 \\(1.68E-218)} &	\makecell[t]{29.7 \\(5.56E-137)} &	\makecell[t]{14.5 \\(9.77E-65)} &	\makecell[t]{7.27 \\(3.44E-27)} &	\makecell[t]{95.5 \\(0)} \\
\multirow{2}{*}{80}&	\makecell[t]{15 \\(5.70E-89)} &	\makecell[t]{46.3 \\(2.41E-275)} &	\makecell[t]{33.7 \\(7.94E-184)} &	\makecell[t]{15.4 \\(3.64E-79)} &	\makecell[t]{6.8 \\(2.55E-28)} &	\makecell[t]{101 \\(0)} \\
\multirow{2}{*}{90}&	\makecell[t]{12.6 \\(8.29E-83)} &	\makecell[t]{56.3 \\(0)} &	\makecell[t]{36.6 \\(1.96E-226)} &	\makecell[t]{15.2 \\(1.45E-92)} &	\makecell[t]{6.25 \\(1.21E-29)} &	\makecell[t]{94.7 \\(0)} \\
\multirow{2}{*}{100}&	\makecell[t]{11.2 \\(1.07E-80)} &	\makecell[t]{45.2 \\(0)} &	\makecell[t]{41.5 \\(1.31E-286)} &	\makecell[t]{17.5 \\(7.36E-120)} &	\makecell[t]{5.93 \\(1.19E-29)} &	\makecell[t]{100 \\(0)} \\
\bottomrule
\end{tabular}%
}
\end{table}

\begin{table}[!ht]
\centering
\caption{Statistical F-Test Results: Only Principal eigenmanifold regression (1st).}
\label{tab:F-test_1st}
\resizebox{\textwidth}{!}{%
\begin{tabular}{c | c| c| c| c| c|c}
\hline
Treatment  & CFH2021 & CFH2021 & CFGS2026 &
CFGS2026 & CFGS2026 & CFGS2026
\\ 
 & UD & SD & $\gamma2$ &
$\gamma3.6$ & $\gamma4.4$ & $\gamma6$
\\\hline
 bin\_num & $F_1$ & $F_2$ & $F_3$  & 
 $F_4$ &$F_5$  & $F_6$ \\
 & ($p_1$) & ($p_2$)& ($p_3$) & 
 ($p_4$)&($ p_5$) & ($p_6$)\\
\hline
% ========== MaxIm 部分 ==========
\multirow{2}{*}{10}&	\makecell[t]{27.8 \\(4.09E-06)} &	\makecell[t]{27.4 \\(4.70E-06)} &	\makecell[t]{17.8 \\(1.24E-04)} &	\makecell[t]{9.29 \\(3.93E-03)} &	\makecell[t]{11.2 \\(1.71E-03)} &	\makecell[t]{103 \\(5.60E-13)} \\
\multirow{2}{*}{20}&	\makecell[t]{67.7 \\(3.14E-14)} &	\makecell[t]{62.7 \\(2.02E-13)} &	\makecell[t]{51.4 \\(1.68E-11)} &	\makecell[t]{18.6 \\(2.64E-05)} &	\makecell[t]{9.48 \\(2.38E-03)} &	\makecell[t]{243 \\(1.07E-35)} \\
\multirow{2}{*}{30}&	\makecell[t]{109 \\(6.80E-23)} &	\makecell[t]{115 \\(5.47E-24)} &	\makecell[t]{115 \\(4.90E-24)} &	\makecell[t]{36.2 \\(3.88E-09)} &	\makecell[t]{12 \\(5.83E-04)} &	\makecell[t]{474 \\(1.72E-71)} \\
\multirow{2}{*}{40}&	\makecell[t]{147 \\(4.01E-31)} &	\makecell[t]{162 \\(6.41E-34)} &	\makecell[t]{213 \\(7.81E-43)} &	\makecell[t]{64.7 \\(3.30E-15)} &	\makecell[t]{16 \\(6.96E-05)} &	\makecell[t]{753 \\(1.69E-116)} \\
\multirow{2}{*}{50}&	\makecell[t]{176 \\(1.09E-37)} &	\makecell[t]{216 \\(3.46E-45)} &	\makecell[t]{336 \\(2.03E-66)} &	\makecell[t]{94.6 \\(1.40E-21)} &	\makecell[t]{19.2 \\(1.30E-05)} &	\makecell[t]{1090 \\(1.38E-171)} \\
\multirow{2}{*}{60}&	\makecell[t]{236 \\(4.30E-50)} &	\makecell[t]{255 \\(1.01E-53)} &	\makecell[t]{474 \\(2.78E-93)} &	\makecell[t]{137 \\(1.87E-30)} &	\makecell[t]{26.8 \\(2.53E-07)} &	\makecell[t]{1480 \\(1.60E-235)} \\
\multirow{2}{*}{70}&	\makecell[t]{304 \\(2.79E-64)} &	\makecell[t]{325 \\(3.48E-68)} &	\makecell[t]{633 \\(2.89E-124)} &	\makecell[t]{180 \\(1.38E-39)} &	\makecell[t]{30.6 \\(3.59E-08)} &	\makecell[t]{1840 \\(4.43E-299)} \\
\multirow{2}{*}{80}&	\makecell[t]{283 \\(7.03E-61)} &	\makecell[t]{397 \\(2.41E-83)} &	\makecell[t]{833 \\(8.71E-163)} &	\makecell[t]{238 \\(8.89E-52)} &	\makecell[t]{40.7 \\(2.05E-10)} &	\makecell[t]{2290 \\(0)} \\
\multirow{2}{*}{90}&	\makecell[t]{269 \\(1.68E-58)} &	\makecell[t]{507 \\(7.48E-106)} &	\makecell[t]{1010 \\(3.31E-198)} &	\makecell[t]{235 \\(1.40E-51)} &	\makecell[t]{49.8 \\(1.96E-12)} &	\makecell[t]{2590 \\(0)} \\
\multirow{2}{*}{100}&	\makecell[t]{258 \\(1.30E-56)} &	\makecell[t]{491 \\(7.80E-104)} &	\makecell[t]{1240 \\(2.06E-243)} &	\makecell[t]{319 \\(3.53E-69)} &	\makecell[t]{45 \\(2.15E-11)} &	\makecell[t]{3090 \\(0)} \\
\bottomrule
\end{tabular}%
}
\end{table}

\clearpage
\subsubsection{MLR Results on Weight (Coef.) of Eigenmanifold-1,2,3 \label{app:c1c2c3}}

\paragraph{Variable definition:}~~The following tables report the multiple linear regression (MLR) coefficients for the first three eigenmanifolds in the full model Eq.~\eqref{eq:main}.
The eigenmanifold vectors \(\sigma_k\) are ordered as in Eq.~\eqref{eq:mn_order}, and the experimental manifold vector \(\bar{L}\) follows the ordering in Eq.~\eqref{eq:L_order}.  
For each treatment and discretization level \(N=\texttt{bin\_num}=10,20,\dots,100\), the tables report:
\begin{itemize}
    \item \(q_{\text{eff}} \): effective number of estimated coefficients (number of conjugate eigenmanifold pairs retained in the design matrix);
    \item \(d_f = n - q_{\text{eff}}\): residual degrees of freedom, with \(n = N(N-1)/2\) equal to the number of components of \(\bar{L}\);
    \item \(R^2_{\text{full}}\): coefficient of determination of the full model;
    \item \(c_1, c_2, c_3\): regression coefficients (weight) of the eigenmanifolds corresponding to the eigenvalues with the first, second, and third largest imaginary parts, respectively;  {here $c_1 \equiv c_{\max},\ c_2 \equiv c_{\text{2nd}},\ c_3 \equiv c_{\text{3rd}}$;}
    \item \(t(c_1), t(c_2), t(c_3)\): \(t\)-statistics of \(c_1, c_2, c_3\).
\end{itemize}
The data sources are CFH2021 \cite{dan2021price} and CFGS2026 \cite{dan2026Cyclical}.  
The theoretical definitions of \(\sigma_k\) and \(\bar{L}\) are given in Section~\ref{sec:eigenmanifold} and Eqs.~\eqref{eq:theo_sigma_mn}--\eqref{eq:bar_L_mn}.

\paragraph{Tables of numerical results for coef.}\label{tab:6trt_10N} 
\paragraph{--- CFH2021 UD}
\begin{center}
\begin{tabular}{|c|r|r|r|r|r|r|r|r|r|}
\hline
\multicolumn{1}{|c|}{$N$} & \multicolumn{1}{c|}{$q_{\text{eff}}$} & \multicolumn{1}{c|}{$d_f$} & \multicolumn{1}{c|}{$R^{2}_{\text{full}}$} & \multicolumn{1}{c|}{$c_1$} & \multicolumn{1}{c|}{$c_2$} & \multicolumn{1}{c|}{$c_3$} & \multicolumn{1}{c|}{$t(c_1)$} & \multicolumn{1}{c|}{$t(c_2)$} & \multicolumn{1}{c|}{$t(c_3)$} \\
\hline
10 & 4 & 41 & 0.4230 & \textbf{22.5406} & \textbf{12.0401} & \textbf{3.7198} & 4.47 & 2.33 & 0.67 \\
\hline
20 & 9 & 181 & 0.3510 & \textbf{11.5546} & \textbf{7.0125} & \textbf{3.4419} & 7.47 & 4.51 & 2.14 \\
\hline
30 & 14 & 421 & 0.3282 & \textbf{7.6950} & \textbf{4.8428} & \textbf{2.7820} & 10.41 & 6.56 & 3.70 \\
\hline
40 & 19 & 761 & 0.3059 & \textbf{5.8925} & \textbf{3.7654} & \textbf{2.2565} & 13.06 & 8.36 & 4.96 \\
\hline
50 & 24 & 1201 & 0.2884 & \textbf{4.7155} & \textbf{3.0784} & \textbf{1.8587} & 15.41 & 10.09 & 6.06 \\
\hline
60 & 29 & 1741 & 0.2527 & \textbf{3.9185} & \textbf{2.5655} & \textbf{1.6103} & 16.75 & 11.01 & 6.88 \\
\hline
70 & 34 & 2381 & 0.2502 & \textbf{3.3899} & \textbf{2.2293} & \textbf{1.4226} & 19.28 & 12.73 & 8.10 \\
\hline
80 & 39 & 3121 & 0.2427 & \textbf{2.9700} & \textbf{1.9574} & \textbf{1.2721} & 21.46 & 14.21 & 9.21 \\
\hline
90 & 44 & 3961 & 0.2428 & \textbf{2.5998} & \textbf{1.7249} & \textbf{1.1255} & 23.91 & 15.94 & 10.38 \\
\hline
100 & 49 & 4901 & 0.2115 & \textbf{2.3809} & \textbf{1.5859} & \textbf{1.0380} & 24.25 & 16.23 & 10.61 \\
\hline
\end{tabular}
\end{center}

\paragraph{--- CFH2021 SD}
\begin{center}
\begin{tabular}{|c|r|r|r|r|r|r|r|r|r|}
\hline
\multicolumn{1}{|c|}{$N$} & \multicolumn{1}{c|}{$q_{\text{eff}}$} & \multicolumn{1}{c|}{$d_f$} & \multicolumn{1}{c|}{$R^{2}_{\text{full}}$} & \multicolumn{1}{c|}{$c_1$} & \multicolumn{1}{c|}{$c_2$} & \multicolumn{1}{c|}{$c_3$} & \multicolumn{1}{c|}{$t(c_1)$} & \multicolumn{1}{c|}{$t(c_2)$} & \multicolumn{1}{c|}{$t(c_3)$} \\
\hline
10 & 4 & 41 & 0.8906 & \textbf{21.7755} & \textbf{11.8857} & \textbf{5.2234} & 15.22 & 8.24 & 3.47 \\
\hline
20 & 9 & 181 & 0.8435 & \textbf{11.4586} & \textbf{7.1414} & \textbf{4.1985} & 23.74 & 14.80 & 8.61 \\
\hline
30 & 14 & 421 & 0.7625 & \textbf{7.8068} & \textbf{5.0174} & \textbf{3.1253} & 26.80 & 17.25 & 10.70 \\
\hline
40 & 19 & 761 & 0.6996 & \textbf{5.8583} & \textbf{3.7873} & \textbf{2.3800} & 29.43 & 19.07 & 11.96 \\
\hline
50 & 24 & 1201 & 0.5886 & \textbf{4.7138} & \textbf{3.0537} & \textbf{1.9597} & 28.95 & 18.80 & 12.05 \\
\hline
60 & 29 & 1741 & 0.5766 & \textbf{3.9443} & \textbf{2.5536} & \textbf{1.6527} & 33.54 & 21.76 & 14.08 \\
\hline
70 & 34 & 2381 & 0.5731 & \textbf{3.3780} & \textbf{2.2100} & \textbf{1.4363} & 38.28 & 25.11 & 16.31 \\
\hline
80 & 39 & 3121 & 0.4551 & \textbf{2.9648} & \textbf{1.9247} & \textbf{1.2533} & 34.05 & 22.16 & 14.42 \\
\hline
90 & 44 & 3961 & 0.3681 & \textbf{2.6422} & \textbf{1.7213} & \textbf{1.1259} & 31.75 & 20.73 & 13.56 \\
\hline
100 & 49 & 4901 & 0.3061 & \textbf{2.3764} & \textbf{1.5415} & \textbf{1.0169} & 30.31 & 19.71 & 13.00 \\
\hline
\end{tabular}
\end{center}

\paragraph{--- CFGS2026 gamma=2}
\begin{center}
\begin{tabular}{|c|r|r|r|r|r|r|r|r|r|}
\hline
\multicolumn{1}{|c|}{$N$} & \multicolumn{1}{c|}{$q_{\text{eff}}$} & \multicolumn{1}{c|}{$d_f$} & \multicolumn{1}{c|}{$R^{2}_{\text{full}}$} & \multicolumn{1}{c|}{$c_1$} & \multicolumn{1}{c|}{$c_2$} & \multicolumn{1}{c|}{$c_3$} & \multicolumn{1}{c|}{$t(c_1)$} & \multicolumn{1}{c|}{$t(c_2)$} & \multicolumn{1}{c|}{$t(c_3)$} \\
\hline
10 & 3 & 42 & 0.3626 & \textbf{35.9215} & \textbf{18.8874} & \textbf{-18.9155} & 4.04 & 1.31 & -0.56 \\
\hline
20 & 7 & 183 & 0.2945 & \textbf{17.3027} & \textbf{9.2046} & \textbf{-1.7997} & 6.84 & 2.98 & -0.33 \\
\hline
30 & 11 & 424 & 0.2942 & \textbf{11.6925} & \textbf{5.9276} & \textbf{0.0831} & 10.30 & 4.68 & 0.05 \\
\hline
40 & 14 & 766 & 0.3020 & \textbf{8.9023} & \textbf{4.3772} & \textbf{0.2081} & 14.06 & 6.44 & 0.25 \\
\hline
50 & 19 & 1206 & 0.3073 & \textbf{7.1513} & \textbf{3.5943} & \textbf{0.3981} & 17.63 & 8.43 & 0.80 \\
\hline
60 & 22 & 1748 & 0.3039 & \textbf{5.9918} & \textbf{2.9754} & \textbf{0.3968} & 21.07 & 10.08 & 1.20 \\
\hline
70 & 29 & 2386 & 0.3037 & \textbf{5.1507} & \textbf{2.6120} & \textbf{0.4341} & 24.36 & 11.99 & 1.82 \\
\hline
80 & 34 & 3126 & 0.3064 & \textbf{4.5458} & \textbf{2.3033} & \textbf{0.4075} & 28.02 & 13.86 & 2.27 \\
\hline
90 & 38 & 3967 & 0.2981 & \textbf{4.0424} & \textbf{2.0475} & \textbf{0.4099} & 30.92 & 15.36 & 2.88 \\
\hline
100 & 42 & 4908 & 0.3005 & \textbf{3.6344} & \textbf{1.8510} & \textbf{0.3779} & 34.37 & 17.22 & 3.32 \\
\hline
\end{tabular}
\end{center}

\paragraph{--- CFGS2026 gamma=3.6}
\begin{center}
\begin{tabular}{|c|r|r|r|r|r|r|r|r|r|}
\hline
\multicolumn{1}{|c|}{$N$} & \multicolumn{1}{c|}{$q_{\text{eff}}$} & \multicolumn{1}{c|}{$d_f$} & \multicolumn{1}{c|}{$R^{2}_{\text{full}}$} & \multicolumn{1}{c|}{$c_1$} & \multicolumn{1}{c|}{$c_2$} & \multicolumn{1}{c|}{$c_3$} & \multicolumn{1}{c|}{$t(c_1)$} & \multicolumn{1}{c|}{$t(c_2)$} & \multicolumn{1}{c|}{$t(c_3)$} \\
\hline
10 & 3 & 42 & 0.3020 & \textbf{30.7080} & \textbf{20.4032} & \textbf{0.2845} & 3.50 & 1.84 & 0.01 \\
\hline
20 & 7 & 183 & 0.2176 & \textbf{15.1806} & \textbf{11.3709} & \textbf{1.6187} & 5.16 & 4.24 & 0.41 \\
\hline
30 & 11 & 424 & 0.1934 & \textbf{10.5712} & \textbf{7.4526} & \textbf{1.7007} & 7.12 & 6.16 & 1.18 \\
\hline
40 & 15 & 765 & 0.1838 & \textbf{8.6134} & \textbf{5.5583} & \textbf{1.5516} & 9.28 & 7.81 & 2.04 \\
\hline
50 & 19 & 1206 & 0.1793 & \textbf{7.3550} & \textbf{4.6282} & \textbf{1.6630} & 11.21 & 9.58 & 3.41 \\
\hline
60 & 24 & 1746 & 0.1686 & \textbf{6.7594} & \textbf{3.9316} & \textbf{1.4082} & 13.22 & 10.75 & 3.96 \\
\hline
70 & 29 & 2386 & 0.1687 & \textbf{6.0257} & \textbf{3.4357} & \textbf{1.3313} & 15.12 & 12.34 & 5.05 \\
\hline
80 & 33 & 3127 & 0.1572 & \textbf{5.6944} & \textbf{2.9796} & \textbf{1.1561} & 17.18 & 13.10 & 5.49 \\
\hline
90 & 39 & 3966 & 0.1462 & \textbf{5.0079} & \textbf{2.8177} & \textbf{1.1257} & 17.39 & 14.47 & 6.35 \\
\hline
100 & 43 & 4907 & 0.1473 & \textbf{4.8338} & \textbf{2.6150} & \textbf{1.0768} & 19.75 & 16.01 & 7.35 \\
\hline
\end{tabular}
\end{center}

\paragraph{--- CFGS2026 gamma=4.4}
\begin{center}
\begin{tabular}{|c|r|r|r|r|r|r|r|r|r|}
\hline
\multicolumn{1}{|c|}{$N$} & \multicolumn{1}{c|}{$q_{\text{eff}}$} & \multicolumn{1}{c|}{$d_f$} & \multicolumn{1}{c|}{$R^{2}_{\text{full}}$} & \multicolumn{1}{c|}{$c_1$} & \multicolumn{1}{c|}{$c_2$} & \multicolumn{1}{c|}{$c_3$} & \multicolumn{1}{c|}{$t(c_1)$} & \multicolumn{1}{c|}{$t(c_2)$} & \multicolumn{1}{c|}{$t(c_3)$} \\
\hline
10 & 3 & 42 & 0.6078 & \textbf{25.4990} & \textbf{28.1143} & \textbf{9.5864} & 4.17 & 5.55 & 1.86 \\
\hline
20 & 7 & 183 & 0.4032 & \textbf{8.4594} & \textbf{12.1210} & \textbf{10.0167} & 3.26 & 6.29 & 5.46 \\
\hline
30 & 11 & 424 & 0.2692 & \textbf{5.0739} & \textbf{6.3205} & \textbf{5.5358} & 3.44 & 6.12 & 5.94 \\
\hline
40 & 15 & 765 & 0.2015 & \textbf{3.8204} & \textbf{4.0499} & \textbf{3.5384} & 3.92 & 6.13 & 6.17 \\
\hline
50 & 19 & 1206 & 0.1531 & \textbf{3.0615} & \textbf{2.7657} & \textbf{2.4340} & 4.33 & 5.92 & 6.16 \\
\hline
60 & 23 & 1747 & 0.1221 & \textbf{2.7866} & \textbf{2.0293} & \textbf{1.7263} & 5.17 & 5.79 & 5.95 \\
\hline
70 & 28 & 2387 & 0.0992 & \textbf{2.3889} & \textbf{1.7154} & \textbf{1.3750} & 5.46 & 6.11 & 6.01 \\
\hline
80 & 32 & 3128 & 0.0843 & \textbf{2.2584} & \textbf{1.3792} & \textbf{1.0744} & 6.33 & 6.08 & 5.90 \\
\hline
90 & 38 & 3967 & 0.0747 & \textbf{2.1501} & \textbf{1.1197} & \textbf{0.8732} & 7.10 & 5.86 & 5.75 \\
\hline
100 & 41 & 4909 & 0.0633 & \textbf{1.7518} & \textbf{1.0850} & \textbf{0.7377} & 6.65 & 6.57 & 5.67 \\
\hline
\end{tabular}
\end{center}

\paragraph{--- CFGS2026 gamma=6}
\begin{center}
\begin{tabular}{|c|r|r|r|r|r|r|r|r|r|}
\hline
\multicolumn{1}{|c|}{$N$} & \multicolumn{1}{c|}{$q_{\text{eff}}$} & \multicolumn{1}{c|}{$d_f$} & \multicolumn{1}{c|}{$R^{2}_{\text{full}}$} & \multicolumn{1}{c|}{$c_1$} & \multicolumn{1}{c|}{$c_2$} & \multicolumn{1}{c|}{$c_3$} & \multicolumn{1}{c|}{$t(c_1)$} & \multicolumn{1}{c|}{$t(c_2)$} & \multicolumn{1}{c|}{$t(c_3)$} \\
\hline
10 & 3 & 42 & 0.8250 & \textbf{44.2907} & \textbf{16.3025} & \textbf{1.7949} & 12.55 & 4.62 & 0.48 \\
\hline
20 & 7 & 183 & 0.7364 & \textbf{22.0211} & \textbf{9.6745} & \textbf{3.6257} & 19.05 & 8.49 & 3.09 \\
\hline
30 & 11 & 424 & 0.6881 & \textbf{14.5473} & \textbf{6.2381} & \textbf{2.4419} & 25.64 & 11.18 & 4.34 \\
\hline
40 & 15 & 765 & 0.6438 & \textbf{10.8356} & \textbf{4.4565} & \textbf{1.7320} & 31.35 & 13.11 & 5.09 \\
\hline
50 & 19 & 1206 & 0.6176 & \textbf{8.6928} & \textbf{3.4964} & \textbf{1.3902} & 37.23 & 15.22 & 6.07 \\
\hline
60 & 23 & 1747 & 0.5941 & \textbf{7.2321} & \textbf{2.8388} & \textbf{1.0870} & 42.82 & 17.07 & 6.57 \\
\hline
70 & 27 & 2388 & 0.5644 & \textbf{6.1835} & \textbf{2.3786} & \textbf{0.8483} & 47.26 & 18.46 & 6.62 \\
\hline
80 & 31 & 3129 & 0.5472 & \textbf{5.4117} & \textbf{2.0331} & \textbf{0.7434} & 52.42 & 19.98 & 7.35 \\
\hline
90 & 37 & 3968 & 0.5169 & \textbf{4.8146} & \textbf{1.8184} & \textbf{0.6449} & 55.48 & 21.25 & 7.58 \\
\hline
100 & 41 & 4909 & 0.5048 & \textbf{4.3422} & \textbf{1.6177} & \textbf{0.5728} & 60.34 & 22.79 & 8.11 \\
\hline
\end{tabular}
\end{center}

 \paragraph{Brief summary}\label{pra:sum_c1c2c3}
The following patterns are observed from the tables above.
  \begin{enumerate}
     \item The ordering \(c_1 > c_2 > c_3 > 0\) holds for all \(N=10,\dots,100\) in CFH2021 UD, CFH2021 SD, CFGS2026 \(\gamma=3.6\), and CFGS2026 \(\gamma=6\). It does not hold for \(\gamma=2\) (where \(c_3<0\) at \(N=10,20\)) or for \(\gamma=4.4\) (where \(c_2>c_1\) for small \(N\)).
%     \item The ratio \(c_2/c_1\) is relatively stable across \(N\) within each treatment. The absolute values of the coefficients decrease with \(N\), because the number and scale of the basis vectors change with \(N\); hence coefficients are not directly comparable across different \(N\).
 % \item The \(t\)-statistics agree with the nested \(F\)-tests: \(t(c_k)^2 = F_{\text{deletion}}\) for the corresponding manifold. For example, for CFGS2026 \(\gamma=2\), \(N=10\), \(t(c_1)^2=16.3\) matches the nested \(F(\sigma_{\max})=16.28\) in Table~\ref{tab:F-test_M1st}; \(t(c_2)^2=1.72\) matches \(F(\sigma_{2nd})=1.705\) in Table~\ref{tab:F-test_M2nd}.
   \item The condition number \(\operatorname{cond}(X)\) deteriorates rapidly with \(N\), more severely for CFGS2026 than for CFH2021. The low-order coefficients \(c_1,\dots,c_3\) remain stable and interpretable, whereas high-order coefficients become numerically unstable and are not reported.
    % \item For \(N\ge 20\), \(c_1\) and \(c_2\) are generally significant (\(|t|>1.96\)) in every treatment. At \(N=10\), \(c_2\) is not significant for \(\gamma=2\) (\(t=1.31\)) and \(\gamma=3.6\) (\(t=1.84\)); \(c_3\) is not significant for CFH2021 UD (\(t=0.67\)).
    \item     The overall fit   $R^2_{\text{full}}$ falls from $0.30$--$0.89$ at $N=10$ to $0.06$--$0.51$ at $N=100$, the largest values occurring for CFH2021 SD and $\gamma=6$ and the smallest for $\gamma=4.4$. The ordering is not an artifact of a high explanatory power: for $N \geq 50$ it holds in every treatment, including $\gamma=4.4$, whose $R^2_{\text{full}}$ is the smallest of all $60$ combinations.
 \end{enumerate}

\clearpage
\newpage

\subsection{The dominant-mode mechanics\label{sec:app_dom}}

This appendix supplies the analytic reason for the weight ordering used in the main
text: why, in the expansion $\bar L=\sum_k c_k\sigma_k$, the eigenmanifold
$\sigma_{\max}$ of the eigenvalue with the largest $|\operatorname{Im}\lambda_k|$ carries
the largest weight --- so that Sections~\ref{sec:sum_result} and \ref{sec:statis_Test_result}  may treat it as the principal term,
and, in the velocity-field construction, as the only term. \\

The ordering is derived below. Two conditions are assumed, and they bear on the amplitude (weight) part of
the argument only, never on the geometric selection of the mode: the driving
spectrum is mode-blind and   damping does not
being specified. Both describe the fluctuation state rather than the games themselves,
and neither adds a constraint to the statistical tests of Section~\ref{sec:results}.\\

The mode selected by the largest complex eigenvalue, $\max (|\Im \lambda_k|)$, dominates because its spatial
geometry --- the long wavelength --- coincides with the low-pass filtering property
of the cumulative utility operator. The largest complex eigenvalue corresponds to the mode of the
longest spatial wavelength in strategy space, i.e.\ the geometric mode of smallest spatial frequency (wave-number)~$k$,   the winding number of its eigenvector is 1, and it is \textbf{one wave going round} mode. So, it is a  spatial low-pass filter of time series.  At it root, it is the cumulative utility function pushes modes of different~$k$ with different strengths. 

\paragraph{(1) Cumulative utility:} It is noticed that, all the utility function in this study cases (CFH2021 \cite{dan2021price}  CFGS2026 \cite{dan2026Cyclical}), CF2003 and CFW2005  \cite{dan2003,dan2005} are cumulative utility function. For the concern of stability and cycle, similar utility function were applied to price dispersion  \cite{BJ83,hopkins2002stability,dan2010gradient,lahkar2011dynamic,paul2017dynamic} which lasting decades, but  the mechanics remain abstract. Here, an ideal model to address the concern. This model provide big picture for understand the observation in this study. \\

Without loss of
generality we expand around an ideal system, e.g.\ $$U(x)=\exp(q-x),$$ where
$q$ is the cumulative of $x$, $q=\int_0^x p(x)\,\mathrm{d}x$, $x\in(0,1]$; one
can show that its Nash equilibrium is the uniform distribution.

\paragraph{(2) Driving force differs across $k$ (algebraic structure) and the observation.}
The utility functions of the continuous-strategy games studied here are all
{cumulative}. Algebraically, the utility gradient (the force that changes the
distribution) is
\begin{equation}
\frac{\partial U_i}{\partial x_j}\propto \mathbf{1}_{i\ge j},
\end{equation}
i.e.\ the utility of strategy $i$ accumulates contributions from strategies
$j\le i$ (not higher than $i$) --- in discrete form the lower-triangular
cumulative matrix $$C=\operatorname{tril}(\mathbf 1) = 
\begin{bmatrix}
1      & 0      & 0      & \cdots & 0 \\
1      & 1      & 0      & \cdots & 0 \\
1      & 1      & 1      & \cdots & 0 \\
\vdots & \vdots & \vdots & \ddots & \vdots \\
1      & 1      & 1      & \cdots & 1
\end{bmatrix}_{N\times N}
$$  The dynamical postulate
(replicator type), denoting the density $p$ at $x$ as $p(x)$,  is $$\mathrm{d}p(x)/\mathrm{d}t \propto p(x) \cdot \partial U/\partial
x;$$ its equilibrium linearization gives the Jacobian $$J=P\,C,$$ where
{$P$ is the mean-zero projection}: it projects the image of the cumulative
operator back onto the feasible tangent space (the constraints of total-mass
conservation and zero mean) \cite{2011Sandholm,dan2016}.
\begin{equation}
P = I_N - \frac{1}{N}\mathbf{1}_N \mathbf{1}_N^\top
% = 
% \begin{bmatrix}
% 1-\frac{1}{N} & -\frac{1}{N} & \cdots & -\frac{1}{N} \\
% -\frac{1}{N} & 1-\frac{1}{N} & \cdots & -\frac{1}{N} \\
% \vdots & \vdots & \ddots & \vdots \\
% -\frac{1}{N} & -\frac{1}{N} & \cdots & 1-\frac{1}{N}
% \end{bmatrix}_{N\times N}.
\end{equation}
$P$ is as essential as $C$ in $J=P\,C$ ($C$ supplies
the low-pass and the dispersion, $P$ enforces the tangent-space constraint). 

\begin{equation}
    J = 
\begin{bmatrix}
0 & -(N-1) & -(N-2) & \cdots & -1 \\
0 & 1 & -(N-2) & \cdots & -1 \\
0 & 1 & 2 & \cdots & -1 \\
\vdots & \vdots & \vdots & \ddots & \vdots \\
0 & 1 & 2 & \cdots & N-1
\end{bmatrix}_{N\times N}
\end{equation}
and three properties can be read off column by column. Read column $j$ as the velocity response to a unit increase of $p_j$ projected back onto the tangent space by $P$, so that total mass is conserved. \\ 

The expression quoted here is the eigenvalue of $P\,C$; the displayed integer matrix is $N\,P\,C$, so $\lambda_k=N\lambda_k(P\,C)=N(\tfrac12-\tfrac{i}{2}\cot\tfrac{\pi k}{N})$, while the Jacobian $J=\tfrac1N P\,C$ has $\lambda_k(J)=\lambda_k(P\,C)/N=\frac{1}{2N}-\frac{i}{2N}\cot\frac{\pi k}{N}$, which is the dispersion used in paragraphs (3)–(4). Frequency ratios, and hence the ordering of the modes, are identical in all three normalizations.\\

\textbf{Eigensystem}:
\begin{itemize}
    \item 

complex eigenvalue:
\[
% \lambda_0 = 0,\qquad 
\lambda_k = \frac{1}{2} - \frac{i}{2}\cot\left(\frac{\pi k}{N}\right),\quad k=1,2,\ldots,N-1.
\]

    \item 
complex eigenvector:
% \[
% v^{(0)} = (1,0,0,\ldots,0)^\top,
% \]
\begin{equation}
v^{(k)} = 
\begin{bmatrix}
\label{eq:eigenvector_k}
1 \\
e^{2\pi i k / N} \\
e^{4\pi i k / N} \\
\vdots \\
e^{2\pi i k (N-1)/N}
\end{bmatrix}
=
\left( e^{2\pi i k (j-1)/N} \right)_{j=1}^{N},
\quad k=1,\ldots,N-1. 
\end{equation} 
    \item 
eigenmanifold $\sigma^k$: 
\begin{equation*}
\sigma^{mn}_k=\frac{\pi}{N}\sin\!\left(\frac{2\pi k(m-n)}{N}\right), \quad N-1\ge n> m\ge 1, 
\end{equation*}
\end{itemize}
\textbf{Travelling wave related observation}:
\begin{itemize}
    \item 
\textbf{travelling wave}:denote   \[
\theta_k \equiv \frac{2\pi k}{N},\qquad
\omega_k=\frac12\cot\frac{\theta_k}{2}=\frac12\cot\frac{\pi k}{N},\qquad \phi_0 \in (-\pi, \pi].
\]
its related observations are:

\begin{eqnarray} 
\text{wave function}& \quad u_j(t)&=e^{t/2}\cos\!\big(\theta_k(j-1)-\omega_k t\big),\\[4pt]
\text{crest function}&  \quad  j_{\text{crest}}(t)&=1+\frac{\omega_k t+\varphi_0}{\theta_k},\\[4pt]
\text{phase velocity} & \quad v_\varphi&=\frac{\omega_k}{\theta_k} 
=\frac{N}{4\pi k}\cot\frac{\pi k}{N},\\[4pt]
\text{wave length} & \quad \lambda_{\text{wave}}&=\frac{N}{k}
=\frac{2\pi}{\theta_k},\\[4pt]
\text{period} & \quad  T&=\frac{2\pi}{\omega_k}
=4\pi\tan\frac{\pi k}{N}
\end{eqnarray}  
\end{itemize}

\textbf{Interpretation}~~(1) These equations illustrate how the observation driven by the utility function.  Meanwhile, these illustrate the dynamics observations are systematic consistency, which can be verified in experiment data.\\

(2) Every local change has a global response. Column $j$ equals $-(N-j+1)$ in the $j-1$ rows above the diagonal and $+(j-1)$ in the remaining $N-j+1$ rows, so the two blocks cancel exactly (column sums are zero). For $N=6$,
\begin{gather*}
\text{raise }p_2\text{ by }\varepsilon:\quad
\delta\dot{x}=\varepsilon\,(-5,\,1,\,1,\,1,\,1,\,1)^{\mathsf T},\\[2pt]
\text{raise }p_6\text{ by }\varepsilon:\quad
\delta\dot{x}=\varepsilon\,(-1,\,-1,\,-1,\,-1,\,-1,\,5)^{\mathsf T}.
\end{gather*}
With $p_2$ it is position $1$ that moves hard downwards and \emph{all} positions $2$--$6$ that move up together, each by the same amount; with $p_6$ it is the five lower positions that move down together and position $6$ that moves up. \\ 

\textbf{Note}~~In this ideal model, \(J=PC\) is diagonalized by the orthogonal discrete Fourier basis. In general games, asymmetric interactions break this structure; the eigenvectors may become non-orthogonal, and cross-mode projections could bias the estimated manifold weights. This is ignored here. 

\paragraph{(3) Spectrum of the cumulative operator.\label{par:spectrum}}
The operator $C$ (and $P\,C$) is diagonalized exactly by the Fourier basis; the
finite-$N$ closed form is
\begin{equation}
\lambda_k(N)=\frac{1}{2N}-\frac{\mathrm{i}}{2N}\cot\frac{\pi k}{N},\qquad
\lvert\operatorname{Im}\lambda_k\rvert=\frac{1}{2N}\cot\frac{\pi k}{N}
\to\frac{1}{2\pi k}\quad(N\to\infty).
\end{equation}
The linearized transfer gain (``spectral response''; the driving) then has two
monotonic facts that run in opposite directions in $k$:
\begin{itemize}
\item {Amplitude (response gain):} $\lvert\lambda_k\rvert=1/(2\pi k)$;
  the smaller $k$ (the longer the wavelength), the larger the gain of $C$ ---
  $C$ is a spatial low-pass filter: long waves are amplified, short waves
  suppressed;
\item {Time frequency:} $\tilde\omega_k=\lvert\operatorname{Im}\lambda_k\rvert
  =1/(2\pi k)$ decreases strictly with $k$, and its maximum sits exactly at $k=1$.
\end{itemize}
Consequently the longest spatial wave ($k=1$) is at the same time the mode pushed
{most strongly} by the cumulative force field is a spatial low-pass filter (largest gain) and the one
rotating {fastest} in time (largest $\lvert\operatorname{Im}\lambda\rvert$);
the criterion ``take the complex eigenvalue with the largest imaginary part'' as
the dominant mode selects exactly this geometric mode --- the two objects are one
and the same, with no separate assumption needed.

\paragraph{(4) Discrete fluctuation--dissipation closure.} That identification concerned the geometry of the force field, not the amplitudes: it fixes which mode is favoured, not how strongly it is excited. The amplitudes are a property of the fluctuation state, and since model  is conservative they must be supplied from outside, by explicit dissipation and noise. Two assumptions supply this excitation model:
\begin{itemize}
\item {(W) white noise:} $\sigma_k^2\equiv\sigma^2$, independent of $k$ --- the
idiosyncratic perturbations carry no preferred spatial scale;
\item {(D) damping does not decrease with $k$:} $\gamma_k$ non-decreasing in $k$,
as satisfied by cumulative smoothing and by spatial diffusion alike.
\end{itemize}
The real part of the discrete eigenvalues, $\operatorname{Re}\lambda_k(N)=1/(2N)$, is
common to all modes and is therefore neglected. Writing the deviation in the discrete
Fourier basis,
$\delta p=\sum_k\operatorname{Re}[c_k\eta_k]$, the modes decouple and the
fluctuation--dissipation balance gives
\begin{equation}
\dot c_k=(\mathrm i\omega_k-\gamma_k)c_k+\sqrt{2\sigma_k^2\gamma_k}\,\xi_k
\qquad\Longrightarrow\qquad
\mathbb E|c_k|^2=\frac{\sigma_k^2}{2\gamma_k}=\frac{\sigma^2}{2\gamma_k},
\end{equation}
so the $k$-dependence of the mode strength comes entirely from the damping spectrum.
The time-averaged angular momentum
$\bar L^{mn}=\sum_t[\delta p_m\dot{\delta p}_n-\delta p_n\dot{\delta p}_m]$ projects onto
the kinematic basis as $\bar L=\sum_kw_k\sigma_k$ with $w_k=\omega_k\mathbb E|c_k|^2$
(the cross-mode terms average out because the $\omega_k$ are distinct), and with
$\omega_k=|\operatorname{Im}\lambda_k|=1/(2\pi k)$ of Eq.~(3)
\begin{equation}\label{eq:1_k}
w_k=\frac{\sigma^2\omega_1}{2}\cdot\frac{1}{k\,\gamma_k}
\qquad\Longrightarrow\qquad
% \arg\max_k w_k=\arg\min_k\big(k\,\gamma_k\big)=1
\frac{w_k}{w_1} =  \frac{1}{k} 
\quad\text{(without specified $\gamma_k$)},
\end{equation}
which is thus
{derived}, at the price of assumptions (W) and (D). 

\paragraph{(5) Observational closure.}
In sum, 
\begin{itemize}
    \item 
Items (1)--(3) show that the cumulative force
field is a spatial low-pass filter: the transfer gain $\lvert\lambda_k\rvert=1/(2\pi k)$
and the time frequency $\tilde\omega_k=\lvert\operatorname{Im}\lambda_k\rvert=1/(2\pi k)$
both attain their maximum at $k=1$. The principal manifold is therefore the
{gravest mode} of the force field --- the longest-wave component, amplified most
and turned fastest --- while fine spatial ripples are simultaneously slow and
suppressed. This is an optical-branch dispersion of a non-local operator, not a
locally coupled medium.

    \item 
    Item~(4) supplies the weights: under a
mode-blind driving (W) and damping that does not decrease with $k$ (D), the
fluctuation--dissipation strength $w_k=\omega_k\sigma^2/(2\gamma_k)$ decreases
monotonically in $k$. The measured angular momentum therefore has the unique expansion
$\bar L=\sum_kw_k\sigma_k$ on the kinematic eigenmanifold basis, with its principal
component falling on $k=1$ by itself --- no additional external driving is needed, and
no empirical rule is invoked. 

\end{itemize}
The dominant-mode criterion
$\arg\max_k\lvert\operatorname{Im}\lambda_k\rvert$ is thus the expression of the force
field's {intrinsic mode}: it is a statement about the geometry of accumulation,
and the angular-momentum protocol measures exactly this weight distribution $w_k$.

\paragraph{(6) Winding number}
In a eigenvector, shown in Eq. \ref{eq:eigenvector_k}, $k$ is the {spatial frequency} of the mode (a discrete wavenumber, $k/N$ cycles per strategy bin), i.e.\ the {winding number} of the phase field of the eigenvector around the strategy domain. Since the strategy domain consists of $N$ discrete points, the winding number is a sum over adjacent strategies rather than a contour integral:
\begin{equation}
\kappa^{(k)}\;=\;\frac{1}{2\pi}\sum_{j=1}^{N}\Delta^{(k)}_{j},
\qquad
\Delta^{(k)}_{j}\;=\;\mathrm{pv}\!\left(\arg \xi^{(k)}_{j+1}-\arg \xi^{(k)}_{j}\right),
\qquad
\xi^{(k)}_{N+1}\equiv \xi^{(k)}_{1},
\label{eq:winding-discrete-en}
\end{equation}
where $\mathrm{pv}(\cdot)$ denotes the principal value in $(-\pi,\pi]$ and the sum runs over one closed cycle ($N$ adjacent pairs, including the closing pair $N\to 1$). For a closed cycle $\kappa^{(k)}$ is always an integer. For the $k$-th Fourier mode $\xi^{(k)}_{j}=\omega^{k(j-1)}$ with $\omega=e^{2\pi i/N}$, every increment is the same,
\begin{equation}
\Delta^{(k)}_{j}\;\equiv\;\frac{2\pi k}{N}\quad(\forall j)
\qquad\Longrightarrow\qquad
\kappa^{(k)}=k ,
\label{eq:winding-fourier-en}
\end{equation}
that is, the accumulated phase change over one traversal is exactly $k$ times $2\pi$, and the phase difference between adjacent strategies is the constant $2\pi k/N$. 

\paragraph{(7) Dominant-mode in discrete strategies game}
For the dominant eigenmodes of high dimensional games with discrete strategies, as shown in Table~1 of the main text, we propose: \\

In the original game matrix, the strategy is unordered unlike continuous strategy game have its scales. But from the original game matrix, via the dynamical equations, one can obtain eigenvalues and eigenvectors. Sorting by the absolute value of the imaginary part of the eigenvalues in descending order, the wave numbers of the corresponding eigenvectors are also integers; hence they can be ordered by the winding number of the eigenvectors. For example, once the order around a single ring is fixed, the \(k=2\) mode is its ``halving'': the real parts in the upper and lower halves have the same sign within each half, and the sums of the imaginary parts are negatives of each other. This eigenvector structure is subject to the constraint that the column sums of the Jacobian are zero. After ordering by winding number, the Jacobian should reflect the spatially dependent and long-range driven dynamics of the gradient of the utility function. \\

We hope the propose above could provide an economics-based, utility-driven mechanistic explanation for the dominant modes in discrete strategies, as we have done in continuous strategy game in this paper. This hypothesis adopts a wave-dynamics narrative; the above assertions still require further proof.
\clearpage

% \subsection{专业术语词汇表\label{sec:eigen_keyword}}
\subsection{Table of the Keywords}\label{sec:eigen_keyword}

\begin{longtable}{|p{2.2cm}|p{1.5cm}|p{9.5cm}|}

\hline
\textbf{Keywords} & \textbf{Symbol} & \textbf{Definition (with section and equation references)} \\
\hline
\endfirsthead
\hline
\multicolumn{3}{|c|}{Continued} \\
\hline
\textbf{Chinese} & \textbf{Symbol} & \textbf{Definition (with section and equation references)} \\
\hline
\endhead
\hline
\endfoot
\endlastfoot

Eigenmanifold (Theoretical Eigenmanifold Vector) & $\sigma_k$ & A vector derived from the complex eigenvectors of the Jacobian matrix near the Nash equilibrium. Each eigenvector $\eta_k$ corresponds to one eigenmanifold $\sigma_k$. Defined in Appendix~\ref{sec:eigenmanifold} (Eq.~\ref{eq:mn_order}). \\
\hline
Eigencycle & $\sigma^{mn}$ & A scalar element of an eigenmanifold vector on the two-dimensional subspace $\Omega^{mn}$. It is computed from the modulus and phase of the two components $(m,n)$ of a given eigenvector $\eta_k$. Defined in Appendix~\ref{sec:eigenmanifold} (Eq.~\ref{eq:theo_sigma_mn}). \\
\hline
Experimental Manifold Vector & $\bar{L}$ & A vector measured from human experimental data; its components are the time-averaged experimental subspace angular momenta $\bar{L}^{mn}$ arranged in a fixed order. Defined in Appendix~\ref{sec:exp_L} (Eqs.~\ref{eq:L_order} and \ref{eq:bar_L_mn}). \\
\hline
Experimental Subspace Angular Momentum & $\bar{L}^{mn}$ & A scalar measuring the instantaneous cyclic intensity on the two-dimensional subspace $(m,n)$, computed from the two state variables $p_m(t)$ and $p_n(t)$. Its time average $\bar{L}^{mn}$ forms a component of the experimental manifold vector. Measurement in Appendix~\ref{sec:exp_L} (Eqs.~\ref{eq:experiment_sigma_mn_L} and \ref{eq:bar_L_mn}). \\
\hline
Order of Manifold Vector Components & $\mathcal{O}$ & The fixed rule that arranges the components (scalars of each two-dimensional subspace) of both the theoretical eigenmanifold vector $\sigma_k$ and the experimental manifold vector $\bar{L}$: first fix $i$ from $1$ to $N-1$, then for each $i$ let $j$ run from $i+1$ to $N$, and sequentially place $\sigma^{i,j}$ (or $L^{i,j}$). This order ensures a unified description of the vector space. Defined in Appendix~\ref{sec:eigenmanifold} (Eq.~\ref{eq:mn_order}). \\
\hline
Matrix Representation of Manifold Vector & $\boldsymbol{T}$ & An antisymmetric matrix representation of the experimental or theoretical manifold vector. Rows $i$ and columns $j$ correspond to discrete strategies, and the entry $\sigma_{ij}$ denotes the net flow from strategy $i$ to $j$, satisfying $\sigma_{ji}=-\sigma_{ij}$. This representation visualises net probability transfers. See Fig.~\ref{fig:heatmap_g2_g6_exp} and its description. \\
\hline
Velocity Field Projection & $v(m)$ & A vector field obtained by projecting an eigenmanifold onto the social state space, representing the average net transfer velocity at the strategy-$(m)$ positions, shown in  Eq.~\ref{eq:velocity}. If the position vector is presented in 2d, the velocity is 2d. \\
% \hline
% Eigenmanifold Vector Space & $\mathcal{V}$ & The abstract space spanned by all theoretical eigenmanifold vectors $\sigma_k$ as basis vectors: $\mathcal{V} = \bigoplus_{k=1}^{N} \operatorname{span}\{\sigma_k\}$. Defined at the end of Appendix~\ref{sec:eigenmanifold}. \\
\hline
Principal (subprincipal) characteristic eigenmanifold & $\sigma_{\text{max}}$ ($\sigma_{\text{2nd}}$) & The eigenmanifold corresponding to the complex eigenvalue with the largest (second largest) absolute imaginary part (i.e., the highest oscillation frequency; why this mode is the dominant one is shown in Appendix~\ref{sec:app_dom}.). Defined in Section\ref{sec:step3}.\\
%~1, ``Step~3''. \\
\hline
\caption{Table of keywords}
\label{tab:eigen_keyword}
\end{longtable}

\bibliography{References}

\end{document}